\documentclass[superscriptaddress,aps,prd,nofootinbib]{revtex4-2}

\usepackage{graphicx}
\usepackage{dcolumn}
\usepackage{hyperref}
\usepackage{ulem}
\usepackage{color}
\usepackage{bm}
\usepackage{siunitx}
\usepackage{braket}
\usepackage{amsfonts,amsmath,amssymb,bm,tensor}
\usepackage{ifpdf}
\usepackage{slashed}
\usepackage[mathscr]{eucal}
\usepackage[utf8]{inputenc}
\usepackage{cancel}

\begin{document}


\title{Strong clustering of primordial black holes from Affleck-Dine mechanism}

\author{Masahiro Kawasaki}
\affiliation{ICRR, University of Tokyo, Kashiwa, 277-8582, Japan}
\affiliation{Kavli IPMU (WPI), UTIAS, University of Tokyo, Kashiwa, 277-8583, Japan}
\author{Kai Murai}
\affiliation{ICRR, University of Tokyo, Kashiwa, 277-8582, Japan}
\affiliation{Kavli IPMU (WPI), UTIAS, University of Tokyo, Kashiwa, 277-8583, Japan}
\author{Hiromasa Nakatsuka}
\affiliation{ICRR, University of Tokyo, Kashiwa, 277-8582, Japan}

\begin{abstract}
Primordial black hole (PBH) is a fascinating candidate for the origin of binary merger events observed by LIGO-Virgo collaboration.
The spatial distribution of PBHs at formation is an important feature to estimate the merger rate.
We investigate the clustering of PBHs formed by Affleck-Dine (AD) baryogenesis, where dense baryon bubbles collapse to form PBHs.
We found that formed PBHs show a strong clustering due to the stochastic dynamics of the AD field.
Including the clustering, we evaluate the merger rate and isocurvature perturbations of PBHs, which show significant deviations from those without clustering.
\end{abstract}

\maketitle

\tableofcontents

\section{Introduction}
\label{sec: Intro}

In recent years, the LIGO-Virgo collaboration has detected gravitational waves emitted by merging binary black holes~\cite{Abbott:2016blz,LIGOScientific:2018mvr,Abbott:2020niy},
which revealed the existence of black holes with masses $\sim \mathcal{O}(10-100) M_\odot$.
Interestingly, many of observed black holes have heavy masses around $30 M_\odot$.
The origin of these massive black holes is still unknown.
One fascinating candidate is the primordial origin~\cite{Bird:2016dcv,Kashlinsky:2016sdv,Sasaki:2016jop,Carr:2016drx,Kawasaki:2016pql,Clesse:2016vqa,Eroshenko:2016hmn,Inomata:2016rbd,Ali-Haimoud:2017rtz,Inomata:2017vxo,Ando:2017veq,Ando:2018nge,Raidal:2018bbj,Liu:2018ess,Vaskonen:2019jpv,Gow:2019pok,Liu:2019rnx,Wu:2020drm,DeLuca:2020bjf,DeLuca:2020qqa}.

Primordial black holes (PBHs) are produced from large density fluctuations in the early Universe~\cite{Hawking:1971ei,Carr:1974nx,Carr:1975qj}.
Such large density fluctuations can be produced by inflation~\cite{GarciaBellido:1996qt,Yokoyama:1995ex,Kawasaki:1997ju} or other mechanisms.
However, if the PBHs have a broad mass spectrum, the corresponding density fluctuation can be severely constrained by cosmological observations.
For example, the observation of the CMB $\mu$-distortion excludes the density fluctuations seeding the PBHs with masses $6 \times 10^4 M_\odot \lesssim M_\mathrm{PBH} \lesssim 5 \times 10^{13} M_\odot$~\cite{Kohri:2014lza}.
In addition, large curvature fluctuations induce the second-order gravitational wave~\cite{Saito:2008jc,Saito:2009jt}, which is constrained by the pulsar timing array experiments.
As a result, the PBHs with masses $0.03 M_\odot \lesssim M_\mathrm{PBH} \lesssim 10 M_\odot$ are severely constrained~\cite{Bugaev:2010bb} (see also~\cite{Carr:2020gox}).

These constraints can be weakened or evaded if the density perturbations seeding the PBHs do not follow the Gaussian distribution,
which is the case, for example, for large density fluctuations with non-Gaussianity~\cite{Garcia-Bellido:2017aan} or nonlinear local objects \cite{Hawking:1987bn,Caldwell:1995fu,Garriga:1992nm,Cotner:2016cvr,Nakama:2016kfq,Deng:2017uwc,Kitajima:2020kig,Kawana:2021tde}.
One of the latter mechanisms is realized by the PBH formation model utilizing the Affleck-Dine (AD) baryogenesis~\cite{Affleck:1984fy,Dine:1995kz}, which is studied in~\cite{Dolgov:1992pu,Dolgov:2008wu,Blinnikov:2016bxu,Hasegawa:2017jtk,Hasegawa:2018yuy,Kawasaki:2019iis}.
While the AD baryogenesis is originally suggested as a model explaining the baryon asymmetry of the Universe in a supersymmetric setup, it is also extended in various contexts.
For example, the AD baryogenesis can work well in the case of large extra dimensions~\cite{Mazumdar:2001nw,Allahverdi:2001dm} and can produce non-thermal dark matter~\cite{Enqvist:1998en,Fujii:2001xp,Fujii:2002kr} through the decay of Q-balls~\cite{Enqvist:1997si,Kusenko:1997si,Kasuya:1999wu,Kasuya:2000wx,Enqvist:2000gq}.
In the PBH formation model utilizing the AD baryogenesis, the IR mode of the AD field diffuses by quantum fluctuations during inflation and has multiple vacua just after inflation.
Then, while the origin of the AD field becomes the true vacuum,
the false vacuum has a non-zero field value.
The inhomogeneity of the field value results in the inhomogeneous baryogenesis, which forms baryon-rich bubbles.
At the QCD phase transition, baryons in the bubbles form massive nucleons and generate density fluctuations.
If the bubbles are large enough, the density fluctuations grow sufficiently and then collapse into PBHs at the horizon reentry.
PBHs generated in this scenario can have masses larger than $\mathcal{O}(10) M_\odot$ and are expected to explain the origin of LIGO-Virgo events~\cite{Hasegawa:2017jtk,Hasegawa:2018yuy}.

In these works, the PBH merger rate has been evaluated under the assumption of the random spatial distributions of PBHs.
However, this scenario can result in a significant clustering of PBHs as pointed out in~\cite{Shinohara:2021psq} for supermassive BHs, and the clustering of PBHs can alter the PBH merger rate.
Since the formation of PBHs or bubbles depends on the diffusion of the IR field,
the time evolution of the IR field on a larger scale can affect the bubble formation on a smaller scale.
Therefore, we have to take into consideration the correlation of the PBH formations in two separate points, which can modify the PBH merger rate.
This mechanism of PBH clustering can be applied to other models of PBH formation~\cite{Nakama:2016kfq,Kitajima:2020kig}, which also utilize the quantum diffusion of the IR field.
The various works also investigate the clustered PBH distribution ~\cite{Rubin:2001yw,Khlopov:2004sc,Chisholm:2005vm,Raidal:2017mfl,Ali-Haimoud:2018dau,Desjacques:2018wuu,Belotsky:2018wph,Ballesteros:2018swv,Bringmann:2018mxj,Ding:2019tjk,DeLuca:2020jug,Atal:2020igj,DeLuca:2021hcf,DeLuca:2021hde}
especially for the non-Gaussian fluctuations~\cite{Tada:2015noa,Young:2015kda,Suyama:2019cst,Young:2019gfc,Suyama:2019cst} while there are no sufficient studies about the clustering of PBHs formed by bubbles.

In this paper, we investigate the quantum diffusion of the AD field with more general initial conditions than the previous works~\cite{Hasegawa:2017jtk,Hasegawa:2018yuy,Kawasaki:2019iis} and formulate the clustering of PBHs through the two-point correlation of the PBH formation in a different way from~\cite{Shinohara:2021psq}.
While they obtain the correlation function of the total PBH number density, we consider the mass distribution of PBHs and obtain the mass dependent correlation function, which is consistent with the result in~\cite{Shinohara:2021psq}.
Then we evaluate the cosmological effect of the PBH clustering on the isocurvature fluctuations and the PBH merger rate in the context of the LIGO-Virgo events.
As a result, we find that the clustering significantly enhances the isocurvature fluctuations and then the PBH abundance can be severely constrained depending on the strength of the PBH clustering.
We also find that, since the PBH clustering in this scenario is strong enough, many of the PBHs form three-body or many-body systems, which results in considerable uncertainty in the estimation of the PBH merger rate.

This paper is organized as follows.
In Sec.~\ref{sec: Affleck-Dine HBB}, we briefly summarize PBH formation from the Affleck-Dine baryogenesis.
In Sec.~\ref{sec: PBH formation and correlation}, we formulate the diffusion of the Affleck-Dine field during inflation and derive the PBH formation rate, mass distribution, and correlation function.
Then we discuss cosmological effects of PBH clustering in Sec.~\ref{sec: cosmological effect}.
Sec.~\ref{sec: discussion} is devoted to the summary and discussion of our results.

\section{High baryon bubbles from the Affleck-Dine baryogenesis}
\label{sec: Affleck-Dine HBB}

In this section, we review the PBH formation in the modified version of the AD baryogenesis based on~\cite{Hasegawa:2017jtk,Hasegawa:2018yuy,Kawasaki:2019iis} and estimate the properties of PBHs formed from baryon-rich regions.
The number density and spatial distribution of PBHs are discussed in Sec.~\ref{sec: PBH formation and correlation}.

In the AD baryogenesis, the AD field is selected as one of the flat directions of the supersymmetric (SUSY) potential, along which the scalar fields have no renormalizable potential term unless SUSY is broken.
When the AD field has a baryon number, a phase rotation of the AD field causes baryogenesis.
We consider the AD field $\phi = \varphi e^{i\theta}$ with the potential of
\begin{equation}
  V(\phi) = \begin{cases}
    (m_{\phi}^2 + c_I H^2) |\phi|^2 + V_{\mathrm{NR}}
    & (\mathrm{during ~ inflation}) \\
    (m_{\phi}^2 - c_M H^2) |\phi|^2 + V_{\mathrm{NR}} + V_\mathrm{T} (\phi)
    & (\mathrm{after ~ inflation})
  \end{cases}
  ,
\end{equation}
where $c_I$ and $c_M$ are dimensionless positive constants,
$m_\phi$ is the soft SUSY breaking mass,
and $V_{\mathrm{NR}}$ is the non-renormalizable contribution given by
\begin{equation}
    V_\mathrm{NR} = \left(
    \lambda a_M \frac{m_{3/2} \phi^n}{n M_{\mathrm{Pl}}^{n-3}}+ \mathrm{h.c.}
    \right)
    + \lambda^2 \frac{|\phi|^{2(n-1)}}{M_{\mathrm{Pl}}^{2(n-3)}},
\end{equation}
where $\lambda$ and $a_M$ are dimensionless constants,
and $M_\mathrm{Pl}$ is the reduced Planck mass.
The integer $n (\geq 4)$ is determined by specifying a flat direction.
Flat directions in the minimal SUSY standard model (MSSM) are summarized in \cite{Dine:1995kz}.
After inflation, the AD field acquires the thermal potential $V_\mathrm{T}$, which is written as
\begin{equation}
  V_\mathrm{T}(\phi)
  =
  \begin{cases}
    c_1  T^2 |\phi|^2
    & (|\phi| \lesssim T)
    \\
    c_2 T^4 \ln\left( \frac{|\phi|^2}{T^2} \right)
    & (|\phi| \gtrsim T)
  \end{cases}
  ,
\end{equation}
where $c_1$ and $c_2$ are $\mathcal{O}(1)$ parameters relevant to the couplings between the AD field and the thermal bath.

During inflation, this potential has the positive Hubble induced mass.
If $c_I \lesssim 1$, the AD field value drifts by quantum fluctuations around the potential minimum at $\phi = 0$.
The time evolution of the AD field is beyond the perturbative method, and we need the stochastic formulation to evaluate the scalar field dynamics, which is discussed in Sec.~\ref{sec: PBH formation and correlation}.

After inflation, the thermal potential overcomes the negative Hubble induced mass near the origin.
Then the potential comes to have the multi-vacuum structure and the AD field rolls down to either of the two vacua depending on whether the field value is larger than the local maximum point $\varphi = \varphi_c$ or not.
As the Hubble parameter decreases, the thermal potential or soft SUSY breaking mass term becomes dominant and the vacuum at $\phi \neq 0$ disappears.
In a Hubble patch where the AD field rolls down to the vacuum at $\phi = 0$, the Affleck-Dine baryogenesis does not work well and almost no baryon number is generated.
On the other hand, in a Hubble patch where the AD field rolls down to the vacuum at $\varphi \neq 0$, the AD field begins to oscillate around the origin at $H \simeq H_\mathrm{osc}$ after the vacuum at $\phi \neq 0$ disappears.
Therefore, the Affleck-Dine baryogenesis works, and the produced baryon to photon ratio is given by
\begin{equation}
    \eta_b \equiv \frac{n_b}{s}
    \simeq \epsilon \frac{T_R m_{3/2}}{H_\mathrm{osc}^2}
    \left(\frac{\varphi_\mathrm{osc}}{M_\mathrm{Pl}}\right)^2,
\end{equation}
where $n_b$ is the produced baryon number density, $s$ is the total entropy density, $T_R$ is the reheating temperature, and $\varphi_\mathrm{osc}$ is the field value when the AD field begins to oscillate.
$\epsilon$ is given by
\begin{equation}
    \epsilon
    =
    \sqrt{\frac{c_M}{n-1}}
    \frac{q_b |a_M| \sin[ n\theta_0 + \arg (a_M)]}
    {3\left( \frac{n-4}{n-2}+1 \right)},
\end{equation}
where $q_b$ is the baryon charge of the AD field and $\theta_0$ is the initial phase of the AD field.
In this way, the AD baryogenesis takes place only in the regions with $\varphi > \varphi_c$ at the end of inflation and
the baryon-rich regions are formed.
Hereafter, we call them high-baryon bubbles (HBBs).

Just after inflation, the energy density inside and outside the HBBs is almost the same since the inflaton dominates the total energy during inflation.
Until the QCD phase transition, quarks, which carry the produced baryon number, remain relativistic and the density fluctuation is not generated.
After the QCD phase transition, the baryon number is carried by massive nucleons, which behave as non-relativistic matter.
Therefore, the density contrast between inside and outside the HBBs grows as
\begin{equation}
    \delta
    \equiv
    \frac{\rho^\mathrm{in} - \rho^\mathrm{out}}{\rho^\mathrm{out}}
    \simeq
    \frac{n_b m_b}{\pi^2g_* T^4/30}
    \simeq
    0.3 \eta_b \left(\frac{T}{200\,\mathrm{MeV}}\right)^{-1}
    \theta(T_\mathrm{QCD}-T),
    \label{eq: density contrast}
\end{equation}
where $m_b \simeq 938 \, \mathrm{MeV}$ is the nucleon mass, $\theta(x)$ is the Heaviside theta function, and $T_\mathrm{QCD}$ is the cosmic temperature at the QCD phase transition.

If this density contrast is large enough, the HBBs collapse into PBHs after they reenter the horizon.
For the perfect fluid with the equation of state $p = w\rho$,
the threshold value of the density contrast for the PBH formation is estimated as~\cite{Harada:2013epa}
\begin{equation}
    \delta_c
    \simeq
    \sin^2 \left( \frac{\pi\sqrt{w}}{1+3w} \right).
\end{equation}
Since, $w$ is written in terms of $\delta$ as
\begin{equation}
    w
    =
    \frac{p^\mathrm{in}}{\rho^\mathrm{in}}
    \simeq \frac{p^\mathrm{out}}{\rho^\mathrm{in}}
    =
    \frac{1}{3(1+\delta)},
\end{equation}
the condition for the PBH formation is given by
\begin{equation}
    \delta
    \gtrsim
    \sin^2
    \left(
        \sqrt{ \frac{1+\delta}{3} } \frac{\pi}{2+\delta}
    \right)
    \Longleftrightarrow
    \delta \gtrsim 0.60.
\end{equation}
This condition gives the upper bound of the temperature at the horizon reentry for the PBH formation as
\begin{equation}
     T_c \simeq \min [100\eta_b \, \mathrm{MeV},T_\mathrm{QCD}]
      \geq T,
\end{equation}
where we used Eq.~\eqref{eq: density contrast}.

The temperature at the horizon reentry is related to the produced PBH mass.
We denote the length scale of the HBBs as $2\pi/k$.
When the mode with the wavenumber $k$ reenter the horizon, the whole energy within the Hubble horizon (horizon mass) is given by
\begin{equation}
    M_H(k)
    \simeq
    20.5 M_{\odot}
    \left( \frac{g_*}{10.75} \right)^{-1/6}
    \left( \frac{k}{10^6 \, \mathrm{Mpc}^{-1}} \right)^{-2},
    \label{eq: horizon mass vs wavenumber}
\end{equation}
where $M_{\odot}$ is the solar mass and $g_*$ is the effective degree of freedom of relativistic particles.
The scale $k$ can be also related to the cosmic temperature when such a scale reenter the horizon as
\begin{equation}
    T(k)
    \simeq
    85.5 \,\mathrm{MeV}
    \left( \frac{g_*}{10.75} \right)^{-1/6}
    \left( \frac{k}{10^6 \, \mathrm{Mpc}^{-1}} \right).
\end{equation}
By using this relation, we can relate the horizon mass and the temperature as
\begin{equation}
    M_H(T)
    \simeq
    3.75 M_{\odot}
    \left( \frac{g_*}{10.75} \right)^{-1/4}
    \left( \frac{T}{200 \, \mathrm{MeV}} \right)^{-2}.
\end{equation}
Since the PBH mass roughly corresponds to the horizon mass at the horizon reentry $M_\mathrm{PBH} \sim M_H$, the upper bound of temperature can be translated into the lower limit of the PBH mass as
\begin{equation}
    M \geq M_c
    \simeq
    \max \left[
        15.0 \, \eta_b^{-2},3.75\left( \frac{T_\mathrm{QCD}}{200\,\mathrm{MeV}}\right)^{-2}
    \right]
    \times M_\mathrm{\odot} \left( \frac{g_*}{10.75} \right)^{-1/4}.
    \label{eq: minimum PBH mass}
\end{equation}

In this paper, we focus on PBHs with masses about $30M_\odot$.
For convenience, we define the $e$-foldings during inflation $N_k$ when the scale $k$ exits the horizon as
\begin{equation}
	N_k
	\equiv
	\ln \left(  \frac{k}{H_I a_i} \right)
	=
	\ln\left(  \frac{k}{k_0} \right),
	\label{eq: e-foldings vs wavenumber}
\end{equation}
where $H_I$ is the Hubble parameter during inflation and $a_i$ is the scale factor when the current horizon scale $k_{0} = 2.24 \times 10^{-4} \, \mathrm{Mpc}^{-1}$ exits the horizon.
The PBH formation scale corresponding to the PBH mass $30M_\odot$ is given by
\begin{align}
    N_{\rm PBH} \equiv N_{k_{\rm PBH}} \simeq 22
    \quad,\quad
    k_\mathrm{PBH} \simeq 8.2 \times 10^5~\mathrm{Mpc}^{-1}.
\end{align}

We comment on the bubbles that do not collapse into PBHs.
The baryon number in them can contribute to the baryon asymmetry of the Universe.
However, in order not to spoil the success of the standard BBN, the baryon asymmetry should be sufficiently homogeneous.
Therefore we should require that the spatially averaged baryon density due to the HBBs $\eta_b^B$ should be much smaller than the observed baryon density of the Universe $\eta_b^\mathrm{obs} \sim 10^{-10}$.
However the baryon charge produced in the AD mechanism is determined by the initial phase of the AD field, which takes a random value after the quantum diffusion.
Then the cancellation of the positive and negative baryon charge will relax this constraint\footnote{Even if the cancellation of baryon and anti-baryon does not work well for some reasons, we can evade the constraint from the baryon overproduction by utilizing, for example, the double inflation or L-balls~\cite{Kawasaki:2021xxx}.}.

Note that, since the inhomogeneous contribution of the bubbles to the baryon asymmetry should be negligible, we need another baryogenesis mechanism in order to explain the baryon asymmetry in the Universe.
In the most simple scenario, the required baryon asymmetry is generated by utilizing another flat direction of the MSSM potential.

\section{PBH formation rate and correlation}
\label{sec: PBH formation and correlation}

In this section, we consider the stochastic time evolution of a complex scalar field $\phi$ during inflation and the bubble formation after inflation.
We investigate the correlation between the field values at different points, $\phi(\bm x)$ and $\phi(\bm y)$.
When two points lie within the same Hubble patch, they have a similar field value $\phi(\bm x) \sim\phi(\bm y)$.
After two points are separated by the inflation and lie in different Hubble patches, they start to evolve in different ways, $\phi(\bm x) \neq\phi(\bm y)$.
Thus, the field values have large correlations when they are spatially close.

Once the scalar field exceeds the threshold value of the potential in some domains, such domains finally form HBBs, which can gravitationally collapse to form PBHs later.
We derive the mass spectrum of PBHs and their spatial correlations in the following.

\subsection{Stochastic dynamics of scalar field}
\label{subsec_scalar_field_dynamics}

During inflation, the scalar field acquires fluctuations with amplitude of the Hubble scale at the horizon exit, and we call the fluctuations as IR modes after they exit the horizon.
The dynamics of IR modes of the scalar field during inflation is described by the Langevin equation including the Gaussian noise originating from the quantum fluctuation~\cite{Vilenkin:1982wt,Starobinsky:1982ee,Linde:1982uu}.
The probability distribution function of the scalar field $P(N,\phi)$ follows the Fokker-Planck equation:
\begin{equation}
    \frac{\partial P(N,\phi)}{\partial N}
    =
    \sum_{i = 1,2} \frac{\partial}{\partial \phi_i}
    \left[
        \frac{\partial V(\phi)}{\partial \phi_i} \frac{P(N,\phi)}{3H_I^2}
        +
        \frac{H_I^2}{8\pi^2} \frac{\partial P(N,\phi)}{\partial \phi_i}
    \right],
\end{equation}
where $(\phi_1, \phi_2) \equiv (\mathrm{Re}[\phi],\mathrm{Im}[\phi])$, $V(\phi)$ is the potential of $\phi$, and $N$ is the $e$-foldings during inflation.
The first term on the right-hand side represents the classical force induced by the potential and the second term represents the quantum fluctuation.
As the initial condition, we assume that $\phi$ stays at the origin, $P(N=0,\phi) = \delta^{(2)} (\phi)$.

In our analysis, we approximate that the Hubble parameter is constant during inflation and we neglect the energy density of the AD field.
During inflation, the potential of $\phi$ is dominated by the Hubble induced positive mass:
\begin{equation}
    V(\phi) = \frac{1}{2} c_I H_I^2 \left( \phi_1^2 + \phi_2^2 \right).
\end{equation}
Since we are interested in the amplitude of $\phi$,
we rewrite this differential equation using $(\varphi, \theta) = (|\phi|, \mathrm{arg}\phi)$ as
\begin{align}
    \frac{\partial P(N,\phi)}{\partial N}
    =&
    \frac{1}{3H_I^2}
    \left[
        \left(
            \frac{\partial^2 V(\varphi)}{\partial \varphi^2}
            +
            \frac{1}{\varphi} \frac{\partial V(\varphi)}{\partial \varphi}
        \right)
        P(N,\phi)
        +
        \frac{\partial V(\varphi)}{\partial \varphi}
        \frac{\partial P(N,\phi)}{\partial \varphi}
    \right]
    \nonumber\\
    &+
    \frac{H_I^2}{8\pi^2}
    \left(
       \frac{\partial^2 P(N,\phi)}{\partial \varphi^2}
        +
        \frac{1}{\varphi} \frac{\partial P(N,\phi)}{\partial \varphi}
        +
        \frac{1}{\varphi^2} \frac{\partial^2 P(N,\phi)}{\partial \theta^2}
    \right)
    \nonumber\\
    =&
    c_I'
    \left(
        P(N,\phi)
        +
        \frac{\varphi}{2} \frac{\partial P(N,\phi)}{\partial \varphi}
    \right)
    +
    \frac{H_I^2}{8\pi^2}
    \left(
       \frac{\partial^2 P(N,\phi)}{\partial \varphi^2}
        +
        \frac{1}{\varphi} \frac{\partial P(N,\phi)}{\partial \varphi}
        +
        \frac{1}{\varphi^2} \frac{\partial^2 P(N,\phi)}{\partial \theta^2}
    \right),
    \label{eq: massive Fokker-Planck equation}
\end{align}
where $c_I' \equiv 2c_I/3$.

In order to consider the probability distribution of $\varphi$, we integrate the probability distribution $P(N,\phi)$ over the phase, i.e., we consider
\begin{equation}
    \tilde{P}(N,\tilde{\varphi})
    \equiv
    \left( \frac{H_I}{2\pi} \right)^2
    \tilde{\varphi} \int_0^{2\pi} \mathrm{d}\theta \, P(N,\varphi,\theta),
\end{equation}
where we use the dimensionless field $\tilde{\varphi} \equiv 2\pi \varphi/H_I$.
Note that $\tilde{P}(N,\tilde{\varphi})$ is normalized to satisfy
\begin{equation}
    \int_0^{\infty} \mathrm{d}\tilde{\varphi} \, \tilde{P}(N,\tilde{\varphi})
    = 1.
\end{equation}
Then the differential equation that $\tilde{P}(N,\tilde{\varphi})$ satisfies is
\begin{equation}
    \frac{\partial \tilde{P}(N,\tilde{\varphi})}{\partial N}
    =
    \frac{c_I'}{2}
    \left(
        \tilde{P}(N,\tilde{\varphi})
        +
        \tilde{\varphi} \frac{\partial \tilde{P}(N,\tilde{\varphi})}{\partial \tilde{\varphi}}
    \right)
    +
    \frac{1}{2}
    \left(
       \frac{\partial^2 \tilde{P}(N,\tilde{\varphi})}{\partial \tilde{\varphi}^2}
        -
        \frac{1}{\tilde{\varphi}}
        \frac{\partial \tilde{P}(N,\tilde{\varphi})}{\partial \tilde{\varphi}}
        +
        \frac{\tilde{P}(N,\tilde{\varphi})}{\tilde{\varphi}^2}
    \right).
    \label{eq: tildeP DE}
\end{equation}

As discussed in App.~\ref{App: Prob distro in massive}, for the initial condition of $\tilde{P}(N = 0, \tilde{\varphi}) = \delta(\tilde{\varphi} - \tilde{\varphi}_{\mathrm{init}})$, this differential equation has the solution of
\begin{equation}
    \tilde{P}(N, \tilde{\varphi}; \tilde{\varphi}_{\mathrm{init}})
    =
    \frac{\tilde{\varphi}}{\tilde{\sigma}^2(N)} I_0
    \left(
        e^{\frac{-c_I' N}{2}}
        \frac{ \tilde{\varphi}_{\mathrm{init}}\tilde{\varphi}}{\tilde{\sigma}^2(N)}
    \right)
    \exp
    \left[
        -\frac{ \tilde{\varphi}^2 + e^{-c_I' N}\tilde{\varphi}_{\mathrm{init}}^2 }
        { 2 \tilde{\sigma}^2(N) }
    \right],
\end{equation}
where $I_0(z)$ is the modified Bessel function of the first kind of order $0$, and
$\tilde{\sigma}^2(N) \equiv (1 - e^{-c_I' N})/c_I'$
describes the variance of the field value.
Especially, for $\tilde{\varphi}_{\mathrm{init}} = 0$,
\begin{equation}
    \tilde{P}(N, \tilde{\varphi}; 0)
    =
    \frac{\tilde{\varphi}}{\tilde{\sigma}^2(N)}
    \exp
    \left[
        -\frac{\tilde{\varphi}^2 }{2 \tilde{\sigma}^2(N)}
    \right].
\end{equation}

The formation rate of bubbles is derived by using $\tilde{P}$, which leads to the PBH formation rate.
The formation of bubbles is determined by the field value at the end of inflation.
$\tilde{P}$ represents the probability distribution of the field value at the horizon exit.
The AD field follows the classical dynamics after the horizon exit, and it rolls down the potential toward the origin.
Taking this effect into account, the threshold of the AD field value to form a bubble is determined by $\varphi_{c,\mathrm{eff}}(N) \equiv \mathrm{exp}[c_I'(N_\mathrm{end}-N)/2]\varphi_c$.
The size of a bubble is set by the horizon scale when $\varphi$ exceeds the threshold value $\varphi_{c,\mathrm{eff}}$.
Since the regions with $\varphi > \varphi_{c,\mathrm{eff}}(N_{k_x})$ at $N_{k_x}$ become bubbles with scales larger than $2\pi/k_x$ after inflation, the volume fraction of
such regions is obtained by integrating $\tilde{P}$ as
\begin{equation}
    {B_1}(N_{k_x},\bm{x})
    =
    \int_{\tilde{\varphi}_{c,\mathrm{eff}}(N_{k_x})}^{\infty} \mathrm{d} \tilde{\varphi} \,
    \tilde{P}(N_{k_x}, \tilde{\varphi}; 0)
    =
    \exp
    \left[
        -\frac{\tilde{\varphi}_{c,\mathrm{eff}}^2(N_{k_x})}{2 \tilde{\sigma}^2(N_{k_x})}
    \right].
    \label{eq: B1}
\end{equation}
Note that the physical volume fraction is the same as the comoving volume fraction during inflation since each patch follows the same expansion history whether $\varphi > \varphi_{c,\mathrm{eff}}$ or not.
In the same way as in the Press-Schechter formalism,
the volume fraction of the regions that would later become bubbles with $k\sim k_{\rm PBH}$ is obtained by differentiating ${B_1}$ with respect to $N$ as
\begin{equation}
    {\beta_{N,1}}(N_{k_x},\bm{x})
    \equiv
    \frac{\partial {B_1}(N_{k_x},\bm{x})}{\partial N_{k_x}}.
\end{equation}
For simplicity, we set $c_M, c_1 \sim 1$ and then
\begin{equation}
    \tilde{\varphi}_c = 2 \pi \Delta^{1/2},
    \quad
    \Delta \equiv \frac{T_R^2 M_\mathrm{Pl}}{H_I^3},
\end{equation}
where $\Delta\gtrsim 1 $ is required for the potential to have the vacuum at the origin.

Next, we consider the two-point correlation of the bubble formation rate.
As a first step, we consider the field values at two points $x$ and $y$ separated by the distance $L$ corresponding to the scale $k_L\equiv 2\pi/L$.
The IR modes of $\phi$ at these two points experience the same time-evolution until the scale $k_L$ exits the horizon since the two points are included in the same Hubble patch.
We denote the field value at the horizon crossing of the mode $k_L$ as $\tilde\phi_L$.
After that, field values at $x$ and $y$ experience the independent history of the field evolution.
Therefore, the volume fraction that $\tilde{\varphi}$ at $x$ exceeds the threshold value at $N_{k_x}$ and that $\tilde{\varphi}$ at $y$ exceeds the threshold value at $N_{k_y}$ is given by
\begin{align}
    {B_2}(k_x,k_y, L)
    =&
	\int \int \int \mathrm{d} \tilde{\varphi}_L \mathrm{d} \tilde{\varphi}_x \mathrm{d} \tilde{\varphi}_y \,
	\tilde{P}(N_{k_L},\tilde{\varphi}_L; 0)
	\tilde{P}(N_{k_x}-N_{k_L},\tilde{\varphi}_x; \tilde{\varphi}_L)
	\tilde{P}(N_{k_y}-N_{k_L},\tilde{\varphi}_y; \tilde{\varphi}_L)
	\nonumber\\
	&\times
	\theta\left(  \tilde{\varphi}_x - \tilde{\varphi}_{c,\mathrm{eff}}(N_{k_x})  \right)
	\theta\left(  \tilde{\varphi}_y - \tilde{\varphi}_{c,\mathrm{eff}}(N_{k_y})  \right).
	\label{eq: B2}
\end{align}
The formation rate distribution of two bubbles with the scale of $k_x$ and $k_y$ at a distance of $L$ is derived by differentiating ${B_2}$ with respect to $N_{k_x}$ and $N_{k_y}$:
\begin{equation}
    {\beta_{N,2}} (k_x, k_y, L)
    =
    \frac{\mathrm{d^2} {B_2}(k_x, k_y, L)}
    {\mathrm{d}N_{k_x} \mathrm{d}N_{k_y}}.
\end{equation}

The distribution of the bubbles with respect to the $e$-foldings $N$ can be translated to the PBH distribution with respect to the logarithm of the PBH mass, $\ln (M_\mathrm{PBH})$.
Considering the relation $M \propto e^{-2N}$ derived from Eqs.~\eqref{eq: horizon mass vs wavenumber} and~\eqref{eq: e-foldings vs wavenumber} and the minimum mass of the PBHs $M_c$ in Eq.~\eqref{eq: minimum PBH mass},
the single PBH formation rate with respect to $M$ is given by
\begin{equation}
    {\beta_1}(M)
    =
    \frac{1}{2} {\beta_{N,1}}(N_k)
    \theta(M - M_c),
    \label{eq_beta1}
\end{equation}
where $M$ and $k$ are related by Eq.~\eqref{eq: horizon mass vs wavenumber}.
Similarly the formation rate of two PBHs with masses $M_x$ and $M_y$ at a distance of $L$ is given by
\begin{equation}
    {\beta_2}(M_x, M_y, L)
    =
    \frac{1}{4} {\beta_{N,2}} (k_x, k_y, L)
    \theta(M_x - M_c)
    \theta(M_y - M_c),
    \label{eq_beta2}
\end{equation}
where $M_i$ and $k_i$ with $i = x,y$ are related by Eq.~\eqref{eq: horizon mass vs wavenumber}.

\subsection{Mass spectrum of PBHs}
\label{subsec_PBH_mass_spec}

Now, we derived the PBH formation rate of one PBH as $\beta_1(M)$ in Eq.~\eqref{eq_beta1} and two PBHs separated by $L$ as $\beta_2(M_i ,M_j, L)$ in Eq.~\eqref{eq_beta2}.
Note that $\beta_1$ and $\beta_2$ are the volume fraction of the regions which collapse into PBHs.
Since the PBH mass is typically the horizon mass at the PBH formation, the total mass of PBHs per volume is given by $\int \mathrm{d} \ln(M)~\beta_1(M)\rho_\mathrm{total}(M)$, where $\rho_\mathrm{total}(M)$ is the total energy density of the Universe at the formation of PBHs with mass $M$.

The energy and number density of the PBHs are evaluated by using $\beta_1(M)$ as
\begin{align}
    \bar{\rho}
    =
    \int \mathrm{d}\ln M \, \rho_\mathrm{total}(M) {\beta_1}(M)
    \quad,\quad
    \bar{n}
    =
    \int \mathrm{d}\ln M \, \frac{\rho_\mathrm{total}(M) \beta_1(M)}{M}.
\end{align}
The present energy ratio of PBHs to the total dark matter is written as
\begin{equation}
    f_\mathrm{PBH}
    \equiv
    \frac{\Omega_\mathrm{PBH}}{\Omega_c}
    \simeq
    \left. \frac{\bar{\rho}}{\rho_m}\right|_\mathrm{eq}
    \frac{\Omega_m}{\Omega_c}
    =
    \int \mathrm{d} (\ln M)~
    {\beta_1}(M) \frac{T(M)}{T_\mathrm{eq}}
    \frac{\Omega_m}{\Omega_c},
\end{equation}
where $\rho_m$ represents the energy density of matter,
the subscript ``$\mathrm{eq}$'' represents the matter-radiation equality,
$\Omega_c$ and $\Omega_m$ represent the present density parameters of the dark matter and total non-relativistic matter, respectively,
and $T(M)$ represents the temperature at the formation of PBHs with mass $M$.
The mass spectrum of PBHs is defined by $f_{\rm PBH}=\int \mathrm{d} (\ln M)~f(M)$, and given by
\begin{align}
    f(M) =
    {\beta_1}(M) \frac{T(M)}{T_\mathrm{eq}}
    \frac{\Omega_m}{\Omega_c}.
\end{align}

\begin{figure}[t]
	\centering
	\includegraphics[width=.45\textwidth ]{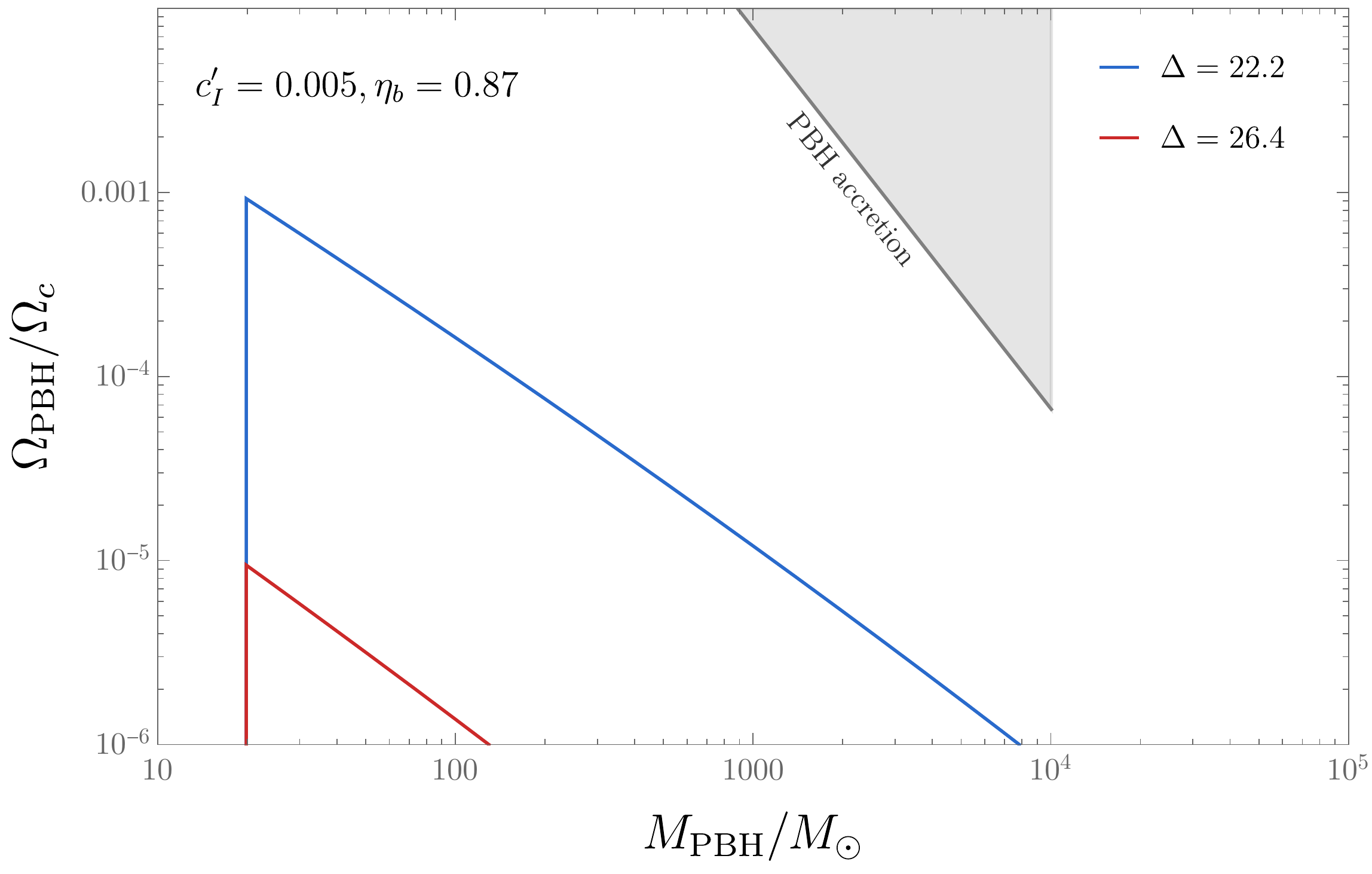}
	\includegraphics[width=.45\textwidth ]{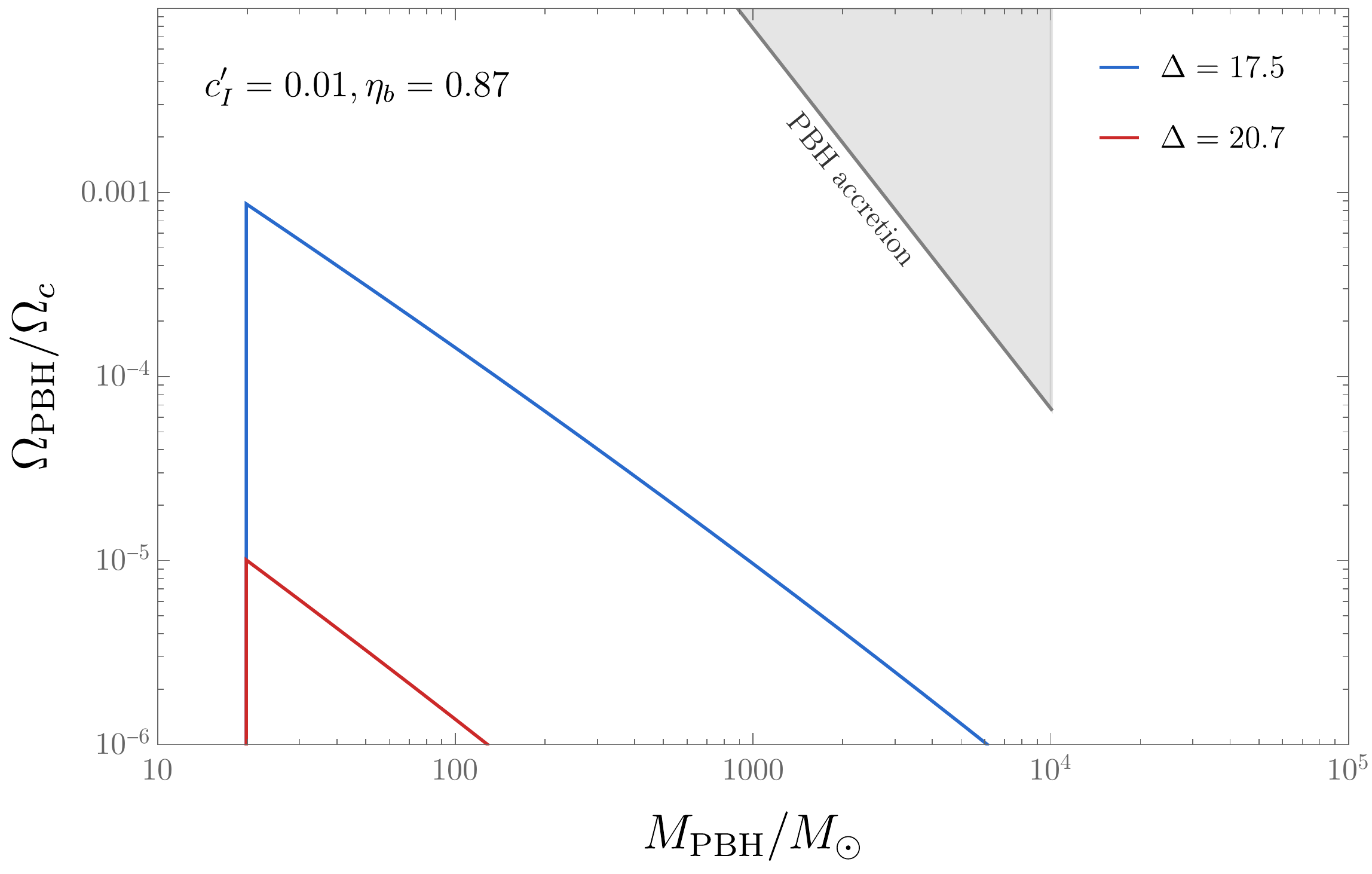}
	\caption{
	PBH mass spectra are plotted for different parameters of the potential.
	The left and right panels represent the results for the different $c_I'$ and the colors of lines represent the different $\Delta$s.
	The total abundance of PBHs is $f_{\rm PBH}\sim 8.6 \times 10^{-4}$ for blue line and $7.9 \times 10^{-6}$ for red line in the left panel, and
	$f_{\rm PBH}\sim 7.8 \times 10^{-4}$ for blue line and $8.2 \times 10^{-6}$ for red line in the right panel.
	The gray region represents the CMB constraint from the PBH accretion under the collisional ionization assumption~\cite{Ali-Haimoud:2016mbv}, which we choose as the most conservative constraint.
	}
	\label{fig_abundance}
\end{figure}

In Fig.~\ref{fig_abundance}, we show the PBH mass spectrum,
where the blue and red lines represent the mass spectrum of our model.
In our model, the relevant constraint on the PBH abundance is given by the PBH accretion~\cite{Ali-Haimoud:2016mbv,Poulin:2017bwe,Serpico:2020ehh}.
Although Ref.~\cite{Serpico:2020ehh} gives the most stringent constraint, we choose the most conservative one~\cite{Ali-Haimoud:2016mbv} as the gray region in Fig.~\ref{fig_abundance}.
The left and right panels represent the different choice of $c_I'$.
We choose $\eta_b$ so that the threshold of the PBH mass comes to $M_\mathrm{c} = 20 M_\odot$.
In our model, the PBH abundance and mass spectrum are controlled by the potential parameters to a certain extent including $f_{\rm PBH}\sim 10^{-3}$, which is a typical value to explain the LIGO-VIRGO events by PBHs assuming that the PBHs are not clustered.

The mass spectrum of PBHs can be approximated by monochromatic distribution when the mass spectrum has a peak shape as in our model.
Using $f_\mathrm{PBH}$, the number density of PBHs with monochromatic mass distribution is given by
\begin{align}
    \bar n
    &\simeq
    \frac{f_{\rm PBH}\Omega_{\rm DM}\rho_{c,0}}{M_{\rm PBH}}
    =
    \left(
    9.65\times 10^{-3}\,{\rm Mpc}
    \right)^{-3}
    ~
    \left(\frac{f_{\rm PBH}}{10^{-3}}\right)
    \left(\frac{30 M_\odot}{M_{\rm PBH}}\right)
    \left(\frac{ \Omega_{\rm DM}}{0.25}\right)
    \left(\frac{ \rho_{c,0}}{0.9\times 10^{-29}\,{\rm g~cm}^{-3}}\right).
\end{align}

\begin{figure}[t]
	\centering
	\includegraphics[width=.45\textwidth ]{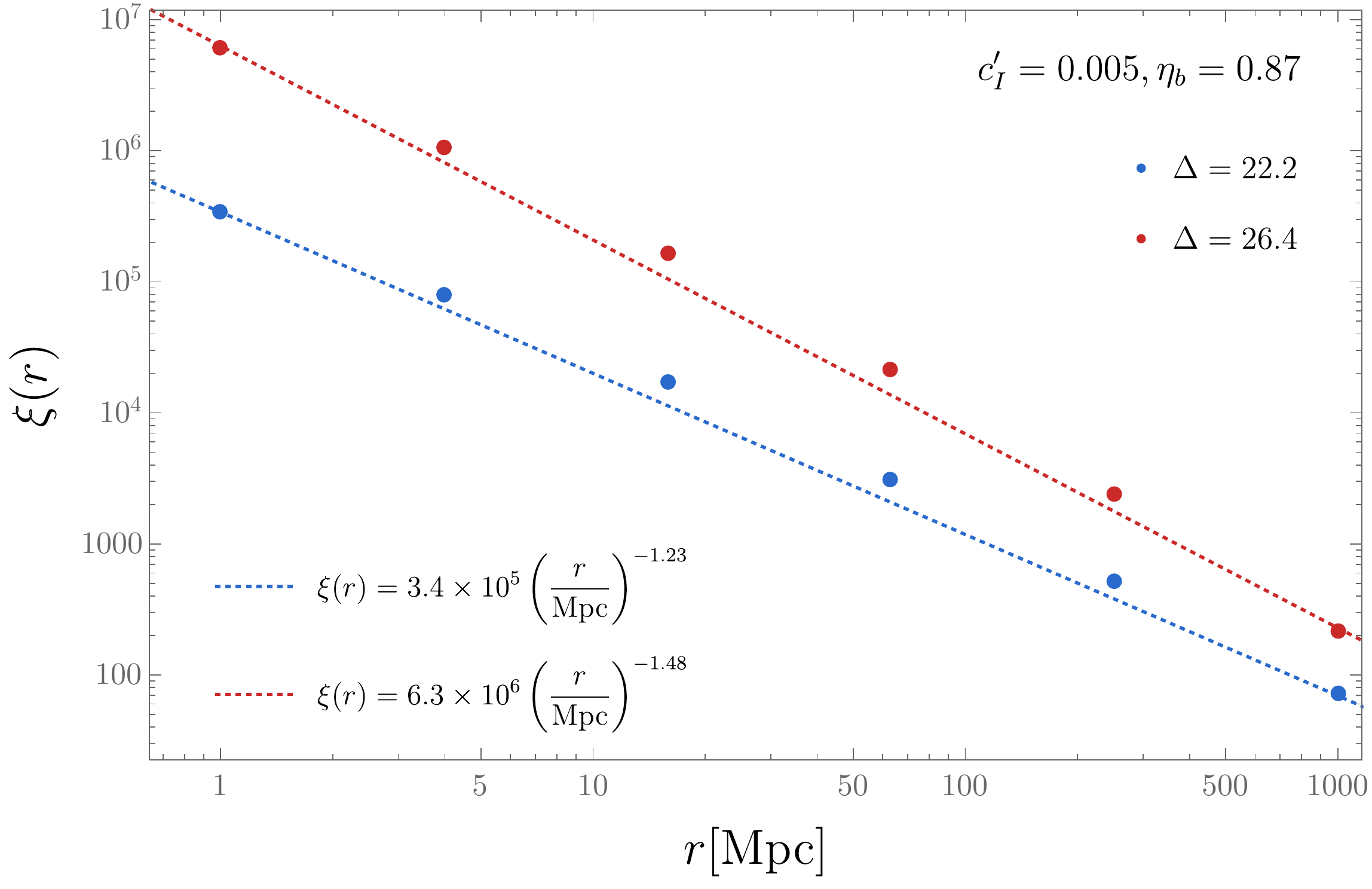}
	\includegraphics[width=.45\textwidth ]{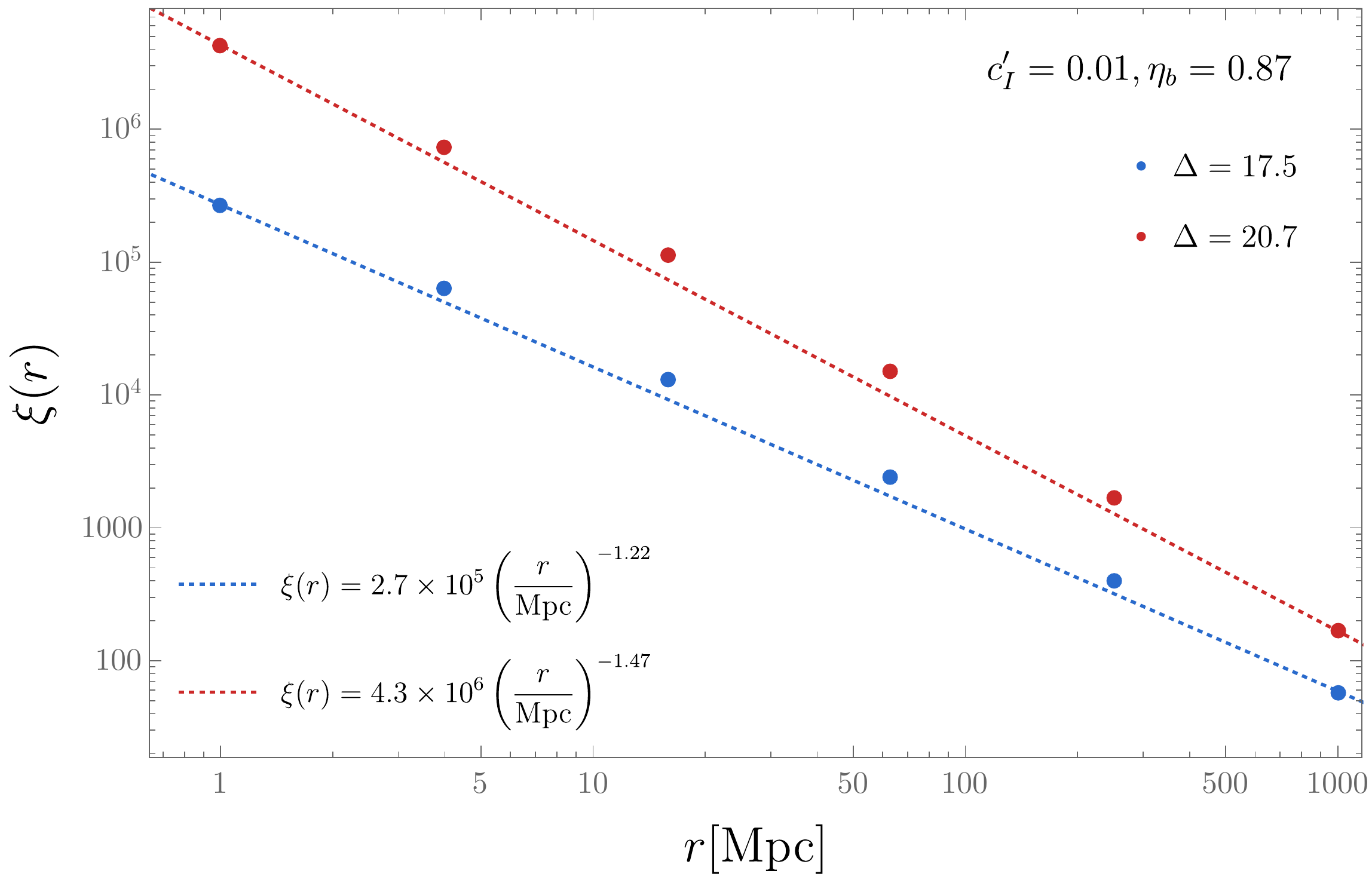}
	\caption{
	The reduced PBH correlation function.
	The parameters are the same as the Fig.~\ref{fig_abundance}.
	The dotted lines are fitting lines of $\xi(r)$ and correspond to the formulae in each panel.
	}
	\label{fig-xi}
\end{figure}

\subsection{Reduced PBH correlation function}
\label{subsec: formulation of PBH correlations}

Here, we formulate the clustering of PBHs using the PBH correlation functions.
When the PBHs are randomly formed independent of the other surrounding PBHs, the PBHs follow the Poisson distribution.
If not, we expect the PBH correlation functions beyond the Poisson distribution.
We derive the PBH correlation functions using the $\beta_2$ in Eq.~\eqref{eq_beta2}.
The outline follows Ref.~\cite{Desjacques:2018wuu} and we extend the formulation for continuous mass spectrum.

First, we start with a finite number of PBHs to calculate their spatial correlation.
The clustering of PBHs is characterized by the two-point correlation function of density perturbations.
We consider $N$ PBHs in a volume $V$, where $i$-th PBH with a mass $M_i$ is located at $\bm x=\bm x_i$ ($i=1,2, \ldots ,N$).
The averaged number density, mass, and energy density of PBHs are given by
\begin{align}
    \bar n = \frac{N}{V}
    \quad,\quad
    \overline{M}
    = \frac{\sum_i M_i}{N}
    \quad,\quad
    \bar \rho = \frac{\sum_i M_i}{V}
    = \bar n \overline{M}.
\end{align}
The density fluctuation of PBHs is written as
\begin{align}
    \delta_\mathrm{PBH}(\bm x)
    \equiv
	\frac{\delta \rho_\mathrm{PBH}(\bm x)}{\bar{\rho}_\mathrm{PBH}}
	=
    \frac{1}{\bar \rho} \sum_{i} M_i\delta^{(3)}(\bm x-\bm x_{i})
    -1.
\end{align}
The two-point correlation function is given by
\begin{align}
& \braket{\delta_\mathrm{PBH}(0)\delta_\mathrm{PBH}(\bm x)}
\nonumber\\
&=
   \Braket{
	    \sum_{i,j}
		\frac{M_i M_j}{\bar \rho^2}
		\delta^{(3)}(\bm x-\bm  x_{i})
		\delta^{(3)}(-\bm  x_{j})
		-
		\frac{
			\sum_{i} M_i \delta^{(3)}(\bm x-\bm x_i)
			+\sum_j M_j \delta^{(3)}(-\bm x_j)
	 	}{\bar \rho}
		+ 1
	}
\nonumber\\
&\equiv
	\frac{\delta^{(3)}(\bm x)}{\bar n} \frac{\overline{M^2}}{( \overline{M} )^2}
	+ \xi(\bm x),
	\label{eq_delta_twopoint}
\end{align}
where the bracket describes the average over the position of PBHs,
$\overline{M^2}\equiv \sum_i M_i^2/N$ is the averaged mass squared,
and $\xi(\bm x)$ is ``a reduced PBH correlation function'' defined by
\begin{align}
    \xi(\bm x)&=
	\sum_{i\neq j } \frac{M_i M_j}{(\bar \rho)^2}
	\Braket{
		\delta^{(3)}(\bm x-\bm  x_i) \delta^{(3)}(-\bm  x_j)
	}
	- 1.
	\label{eq_def_xi_finite}
\end{align}

Next, we extend the definition of the reduced PBH correlation function for continuous mass spectrum of PBHs.
Using the formation rate of two PBHs ${\beta_2}(M_j ,M_J, L)$ in Eq.~\eqref{eq_beta2},
we can deduce the reduced PBH correlation function as
\begin{equation}
    \tilde{\xi}(x)
    \equiv
    \frac{\int \mathrm{d}\ln M_i \mathrm{d}\ln M_j \, \rho_\mathrm{total}(M_i) \rho_\mathrm{total}(M_J) {\beta_2}(M_i, M_j, \bm x)}
    {\left(\int \mathrm{d} \ln M \, \rho_\mathrm{total}(M) {\beta_1}(M)\right)^2}
    - 1.
    \label{eq_def_tildexi}
\end{equation}
Note that $\tilde{\xi}(x)$ vanishes
when there are no correlations in the PBH distribution,
${\beta_2}(M_i, M_j, x) = {\beta_1}(M_i) {\beta_1}(M_j)$.
Here we comment on a subtle difference between $\xi(x)$ and $\tilde{\xi}(x)$.
Strictly speaking, $\tilde{\xi}$ is defined so that $(1+\tilde{\xi}(x))\bar{n}$ represents the expectation value of the PBH number density at a distance $x$ when we assume that one PBH is located at the origin $x = 0$.
Since this assumption increases the PBH formation rate around the origin, the calculation of $\tilde{\xi}$ implicitly assumes the larger number density of PBHs than $\bar n$.
On the other hand, $\xi(x)$ represents the probability that two PBHs form at a distance $x$ when the number density of PBHs is fixed to $\bar n$.
This difference leads to a correction factor to the relation between $\xi(x)$ and $\tilde{\xi}(x)$
We estimate this correction in App.~\ref{app: reduced PBH correlation function} and find that the difference is about $\mathcal O(10\%)$.
Thus, we approximately use $\xi(x)\sim\tilde{\xi}(x)$ in this paper.

We show the reduced PBH correlation function in Fig.~\ref{fig-xi}, where the parameters are the same as in Fig.~\ref{fig_abundance}.
Here we evaluated $\tilde{\xi}(x)$ by assuming the monochromatic mass spectrum with $M=30M_{\odot}$ and ignoring the integration over mass.
As shown in Fig.~\ref{fig-xi}, the decline of $\xi$ becomes milder for smaller $c_I'$.
This is because smaller $c_I'$ leads to a larger variance of $\phi$ during inflation
and the PBH formation on smaller scales is more affected by the fluctuations on larger scales.
In the limit of a scale of the PBH radius,
$\xi$ is proportional to
$\beta_1^{-1}(M_\mathrm{PBH})$, since, in this limit, the AD field follows almost the same evolution in the two distanced points.
(See Eqs.~\eqref{eq: B1} and \eqref{eq: B2}.)
This is why $\xi(x)$ is roughly proportional to $f_\mathrm{PBH}^{-1}$ when we compare the blue and red lines in Figs.~\ref{fig_abundance} and \ref{fig-xi}.

The reduced PBH correlation function around $r\sim 1~$Mpc is fitted by the following formula,
\begin{align}
    \xi(r)
    &=
    \xi_*
    \left(\frac{r}{1~\mathrm{Mpc}}\right)^{-\alpha}
    \label{eq_xi_summary}
    \quad,\quad
    f_\mathrm{PBH} \xi_* \sim (0.3 \, \mathchar`- \, 3) \times 10^{2}
    \quad,\quad
    \alpha\sim 1.2 \, \mathchar`- \, 1.5
\end{align}
where $ k_{\rm PBH} $ is the comoving scale of fluctuations to form PBHs with $M=30M_{\odot}$.
In the following section, we use $\alpha=1.5$ as a typical value and treat $(f_\mathrm{PBH},\, \xi_*)$ as free parameters although they are roughly inversely proportional.

\section{Cosmological effects of PBH clustering}
\label{sec: cosmological effect}

In this section, we discuss the cosmological effects of PBH clustering.
The clustered distribution of PBHs induces the density fluctuation in addition to the Poisson fluctuation as we derived in Eq.~\eqref{eq_delta_twopoint}.
Such additional fluctuations source the isocurvature perturbations.
Moreover, the clustering of PBHs affects the binary formation rate of PBHs, which can modify the merger rate distribution of PBHs.
We discuss these cosmological phenomena in the following.

\subsection{Isocurvature fluctuations}
\label{subsec: isocurvature}

It is known that PBHs induce isocurvature fluctuations due to their Poisson fluctuations, and the clustering also sources isocurvature fluctuations~\cite{Desjacques:2018wuu,Matsubara:2019qzv}.
The power spectrum of PBH density perturbations is defined as
\begin{align}
    P_{\rm PBH}(k)
   & =
    \int
   \mathrm{d} ^3  \bm x ~
   e^{-i\bm k\cdot \bm x}
    \braket{\delta_\mathrm{PBH}(\bm 0)\delta_\mathrm{PBH}(\bm x)}
    =
    P_{\rm Poisson}(k)
    + P_{\xi}(k),
    \\
     P_{\rm Poisson}(k)
     &\equiv
     \frac{1}{\bar n}
     \frac{\overline{M^2}}{(\overline{ M})^2},
     \label{eq_Poisson_power_spec}
      \\
     P_{\xi}(k)
     &\equiv
     \int
     \mathrm{d} ^3  \bm x ~
   e^{-i\bm k\cdot \bm x} \xi(x)
   = 4\pi \int\mathrm{d} x~
   x^2
   \frac{\sin(kx)}{kx} \xi(x),
   \label{eq_Pxi_integral}
\end{align}
where $P_{\rm Poisson}(k)$ comes from the Poisson fluctuation induced by the fluctuation of the number of PBHs, and
$P_{\xi}(k)$ comes from the clustered distribution of PBHs.
When we adapt the power law form of $\xi(r)$ in Eq.~\eqref{eq_xi_summary} with $1<\alpha<3$, $P_{\xi}(k)$ is given by
\begin{align}
    P_{\xi}(k)
     &
   =
  4\pi k^{-3} \xi_*
   (1~\mathrm{Mpc}\times k)^{\alpha}
   \sin \left( \frac{\alpha \pi}{2} \right)
   \Gamma(2-\alpha)
   .
\end{align}
Note that the integration variable $x$ has a lower bound at the scale of the PBH formation and an upper bound at the scale of the observable Universe.

\begin{figure}[t]
	\centering
	\includegraphics[width=.80\textwidth ]{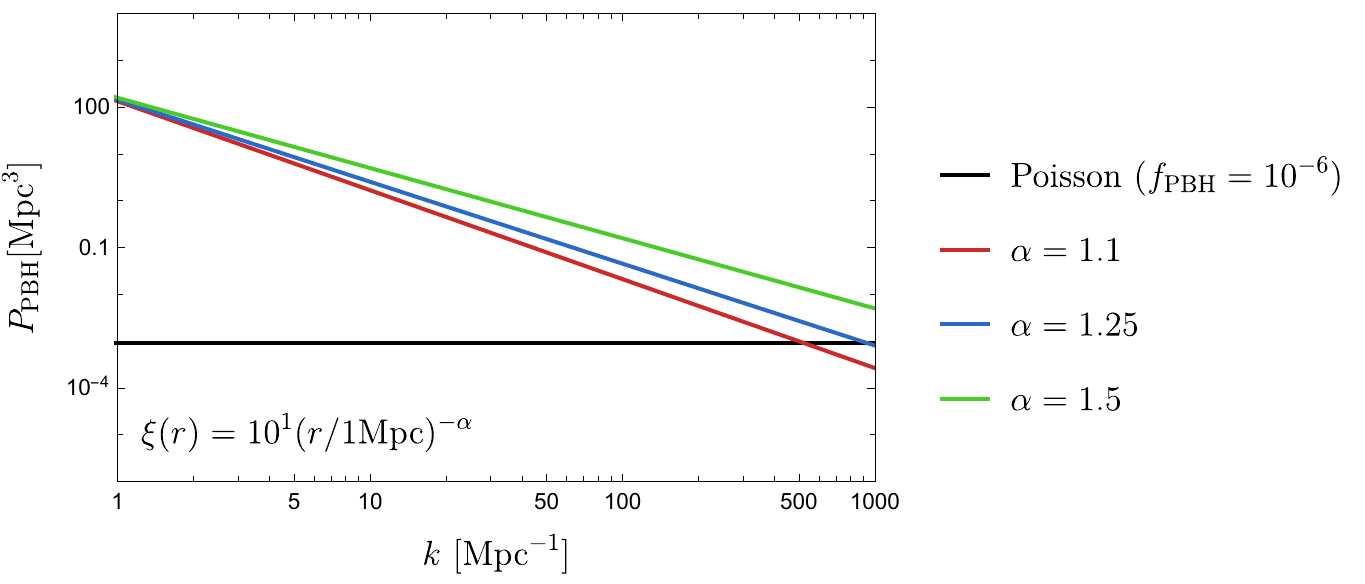}
	\caption{
	The power spectrum of PBH density perturbations $\delta_{\rm PBH}\equiv \delta\rho_{\rm PBH}/\bar\rho_{\rm PBH}$.
	The black line represents the Poisson fluctuation in Eq.~\eqref{eq_Poisson_power_spec} for $f_{\rm PBH} = 10^{-6}$.
	The colored lines represent the power spectrum induced by the clustering of PBHs assuming the reduced PBH function as $\xi(k) = 10^1 (r/1\mathrm{Mpc})^{-\alpha}$.
	The tilts of the colored lines are given by $P_\xi(k)\propto k^{\alpha-3}$.
	}
	\label{fig_isocurvature}
\end{figure}

We compare $P_\xi$ to the Poisson fluctuation in Fig.~\ref{fig_isocurvature}.
The Poisson fluctuation of PBHs (black line) is evaluated for $f_{\rm PBH} = 10^{-6}$.
$P_\xi(k)$ is shown by colored lines, where the reduced PBH function is assumed as $\xi(r) = 10^1 (r/1\mathrm{Mpc})^{-\alpha}$.
The contribution of clustering overcomes the Poisson fluctuation on a larger scale even though the clustering decreases on a larger scale as $\xi(r) \propto  r^{-\alpha}$ with $\alpha<3$.
In all the parameter sets used in Fig.~\ref{fig_abundance}, $P_\zeta$ gives the main contribution to the isocurvature perturbation at $k= \mathcal O(1) \, \mathrm{Mpc}^{-1}$.

\begin{figure}[t]
	\centering
	\includegraphics[width=.80\textwidth ]{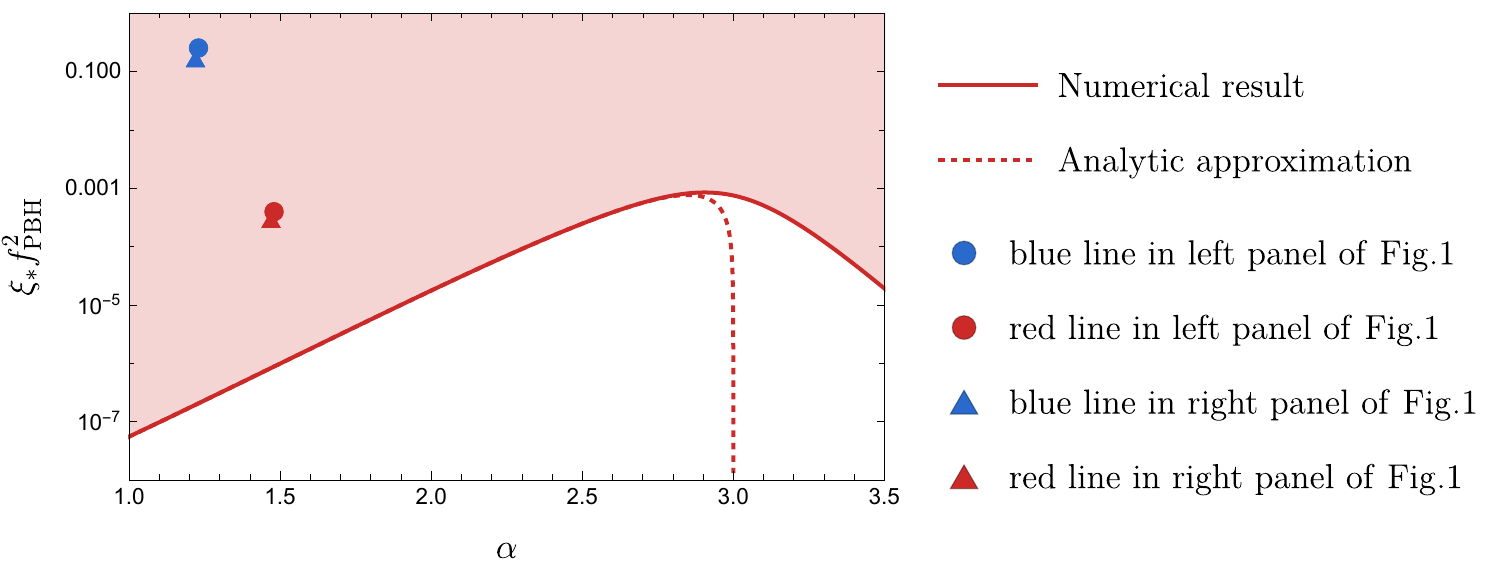}
	\caption{
	The constraint on the PBH abundance and clustering by the CMB observation.
	We approximate the reduced PBH correlation function as $\xi(r) = \xi_*(r/1\mathrm{Mpc})^{-\alpha}$.
	The isocurvature constraint by Planck satellite~\cite{Akrami:2018odb} excludes the strong clustering of PBHs (red region), where the red dotted line represents the analytic formula [Eq.~\eqref{eq_const_isocurvature}] valid for $1<\alpha<3$ and the red solid line represents the numerical result including the lower bound of the integration variable in Eq.~\eqref{eq_Pxi_integral} at $x = 2\pi/k_{\rm PBH}$.
	Each point describes the parameters used in Fig.~\ref{fig_abundance}.
	}
	\label{fig_const_xif2}
\end{figure}

The isocurvature fluctuation is given by
\begin{align}
    P_{\rm iso}(k)
    =
    \left(\frac{\rho_{\rm PBH}}{\rho_{\rm DM}} \right)^2
    P_{\rm PBH}(k)
     =
    f_{\rm PBH}^2
    P_{\rm PBH}(k)
    .
\end{align}
The amount of isocurvature perturbations $\beta_{\rm iso}\equiv P_{\rm iso}/(P_{\rm iso}+ P_{\mathcal R})$ is constrained by CMB observations~\cite{Akrami:2018odb}, where $P_{\mathcal{R}}$ is the power spectrum of curvature perturbations and they assume the flat power spectrum of isocurvature perturbations.
Since our result predicts the blue-tilted power spectrum, we conservatively use the constraint on a large scale, $\beta_{\rm iso}< 0.035 $ for the uncorrelated cold DM case at the scale $k_{\rm low} =0.002\,{\rm Mpc}^{-1}$~\cite{Akrami:2018odb}.
The PBH abundance and clustering are constrained as
\begin{align}
    f_\mathrm{PBH}^2
    \xi_*
    &\lesssim
    1.75\times 10^{-5}
    ~
    \left( 2\times 10^{-3} \right)^{2-\alpha}
    \frac{\pi/2}{\sin \left(\frac{\alpha \pi}{2} \right) \Gamma(2-\alpha)}
    \frac{\beta_{\rm iso}}{0.035}
    \frac{\mathcal P_{\mathcal R}}{2\times 10^{-9}}
    \quad,\quad
    \mathrm{(1<\alpha<3)}.
    \label{eq_const_isocurvature}
\end{align}
We show the constraints on $f_\mathrm{PBH}^2 \xi_*$ in Fig.~\ref{fig_const_xif2},
where the red dotted line represents the analytic formula in Eq.~\eqref{eq_const_isocurvature} and the red solid line represents the numerical result including the lower bound of the integration variable in Eq.~\eqref{eq_Pxi_integral} at $x = 2\pi/k_{\rm PBH}$.
The lower bound of integration affects the results for $\alpha \gtrsim 3 $.
The parameters used in Fig.~\ref{fig_abundance} are also plotted.
Although our model is constrained by the isocurvature perturbations for our parameter sets, we can still avoid it by choosing a smaller abundance of PBHs.
As the abundance is small, the tilt of the clustering, $\alpha$, becomes large, which helps to avoid the isocurvature constraint.

\subsection{Merger rate distribution}
\label{subsec: merger rate calc}

The clustered distribution of PBHs modifies the PBH merger rate distribution~\cite{Raidal:2017mfl,Ding:2019tjk,Ballesteros:2018swv,Bringmann:2018mxj,Young:2019gfc,DeLuca:2020jug,Atal:2020igj}.
Here, we estimate the merger rate distribution with clustering.
For simplicity, we assume the monochromatic PBH mass spectrum with $M_{\rm PBH}=30 M_\odot$.
The PBH merger rate is calculated in two steps.
First, we calculate the probability in which three PBHs (I), (II), and (III) are produced at the comoving distance $r=0$, $x$, and $y$ ($0<x\ll y$).
Second, we estimate when PBHs (I) and (II) form binary and merge.
Since the eccentricity of the binary affects the coalescence time, we need to estimate the angular momentum of the binary induced by PBH (III).
We include the main contribution of angular momentum induced by PBH (III) and neglect the effect of the outer PBHs, which is discussed in \cite{Ali-Haimoud:2017rtz}.
Then, we can derive the current merger rate distribution~\cite{Sasaki:2016jop,Kocsis:2017yty}.

The clustering of PBHs modifies the above discussion through the increased formation rate of PBHs (II) and (III) by factors $\xi(x)$ and $\xi(y)$.
When PBH (II) is more frequently produced, we have more binaries and a larger merger rate.
When PBH (III) is more frequently produced, the binaries tend to acquire larger angular momenta and longer coalescence times.
If PBH (III) is too close to the binary, it may form a three-body system and  disrupt the binary.
Thus, the clustered distribution has two opposite effects on the binary formation rate.
We include those effects in the following calculation to quantitatively estimate the merger rate.

\begin{figure}[t]
	\centering
	\includegraphics[width=.60\textwidth ]{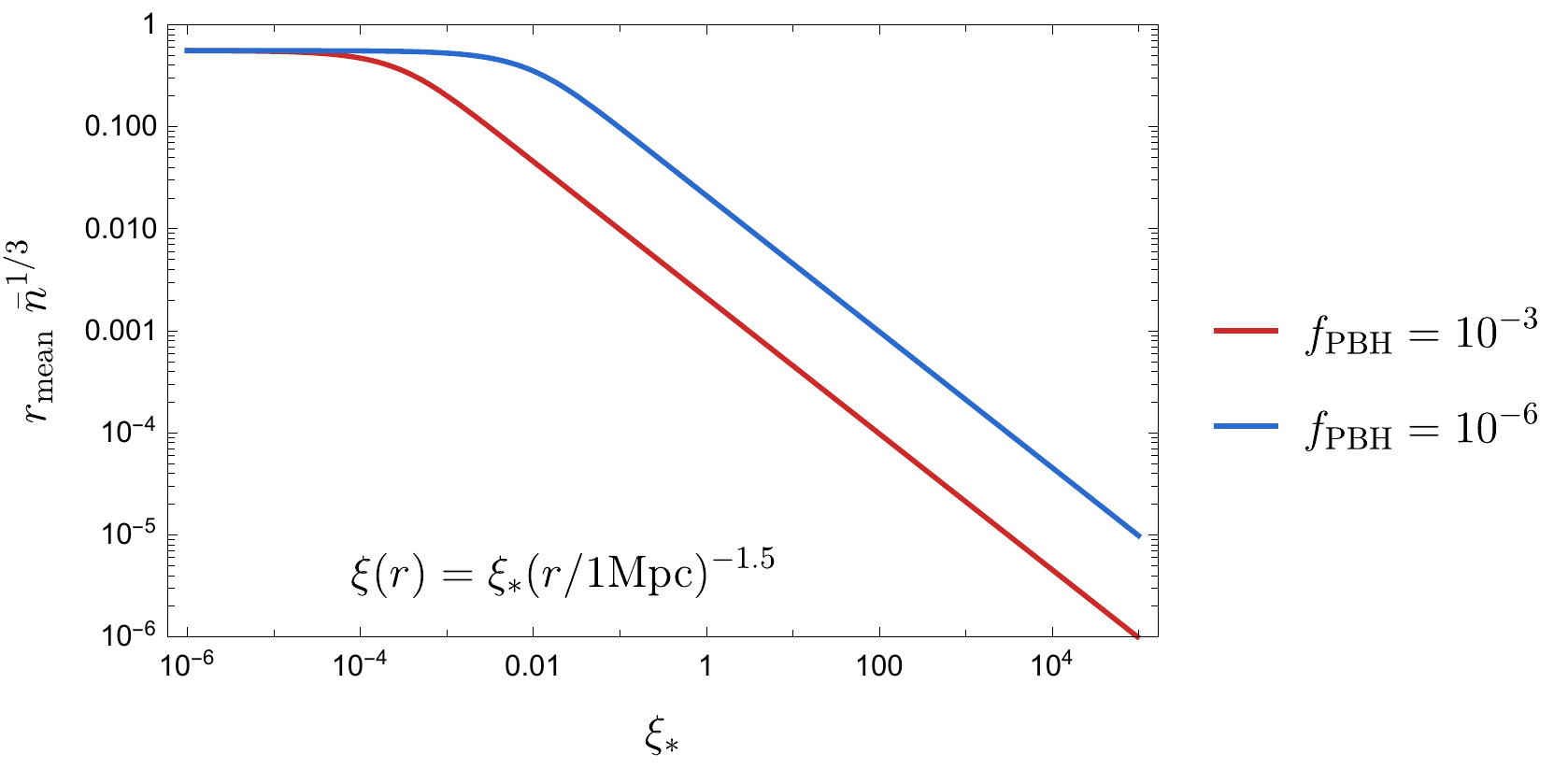}
	\caption{
    The mean separation of the nearest-neighbor PBH [Eq.\eqref{eq_rmean}] with clustering for the PBH abundance $f_{\rm PBH} =10^{-3}$ (red line) and $10^{-6}$ (blue line).
    We assume the reduced PBH correlation function as  $\xi(r)= \xi_*\left(r/1~\mathrm{Mpc}\right)^{-1.5}$.
	}
	\label{fig_typical_Separation}
\end{figure}

First, we investigate the effect of clustering on the distribution of PBHs.
Without clustering, the mean separation length between PBHs is given by $r_{\rm mean} \sim \bar n^{-1/3}$.
With the positive (negative) clustering, $r_{\rm mean}$ becomes smaller (larger).
We estimate $r_{\rm mean}$ between two PBHs, (I) at the origin ($r=0$) and (II) at $r=x$, which is the nearest neighbor of PBH (I).
The PBH formation probability of a PBH at $\bm x$ is given by $(1+\xi(x)) \bar n \mathrm{d}^3 x$.
Since PBH (II) is the nearest neighbor, we require that no other PBH lies between $r\in [0,x]$, which suppress the probability by  $\exp[- \int_0^x \mathrm{d}^3 r~ (1+\xi(r)) \bar n]$.
The probability for the formation of PBH (II) is given by~\cite{Raidal:2017mfl,Ballesteros:2018swv}
\begin{align}
	Q_1(x) &= (1+\xi(x)) ~\bar n ~
	\exp\left[
	-\Gamma(x)
	\right],
	\\
      \Gamma(x) &\equiv
    \frac{ 4\pi \bar n x^3}{3}
		\left(
			1+
			3 x^{-3} \int^x_0\mathrm{d} s \, s^2 \xi(s)
		\right),
\end{align}
where $Q_1(x)$ is normalized as $\int_0^\infty \mathrm{d}^3 x~Q_1(x)=1$.
Using this probability, we define the mean separation of the nearest-neighbor PBHs as
\begin{align}
    r_{\rm mean}
    &\equiv
    \int \mathrm{d} ^3 \bm r ~
    r Q_1(r).
    \label{eq_rmean}
\end{align}
Without clustering, we can analytically calculate Eq.~\eqref{eq_rmean} as $r_{\rm mean}\to \Gamma(1/3)(36\pi)^{-1/3}\bar n^{-1/3}\sim 0.6\bar n^{-1/3}$, which is consistent to the naive estimation, $r_{\rm near}\sim\bar n^{-1/3}$.
We show the normalized $r_{\rm mean}$ in Fig.~\ref{fig_typical_Separation}, where we assume that the reduced PBH correlation function as $\xi(r)= \xi_*\left({r}/1~\mathrm{Mpc}\right)^{-1.5}$.
The horizontal axis represents the strength of clustering.
The clustering of PBHs shortens the mean separation of the nearest-neighbor PBHs, especially for the larger PBH abundance (red line) rather than the smaller PBH abundance (blue line).

Next, we define the probability distribution in which PBH (II) is formed between $x$ and $x+\mathrm{d} x$ and PBH (III) is formed between $y$ and $y+\mathrm{d} y$ as $Q_{12}(x,y) \mathrm{d} ^3 x\mathrm{d} ^3 y$.
First, we focus on the conditional probability $Q_{2}(y|x)$ where PBH (III) is formed at $y$ on the condition that PBH (II) lies at $x$.
Since PBH (III) is the next-to-nearest neighbor of PBH (I), we require that no other PBH lies between $r\in[x,y]$, which leads to $Q_{2}(y|x) =  (1+\xi(y)) \bar n\exp[- \int_x^y \mathrm{d}^3 r~ (1+\xi(r)) \bar n]\theta(y-x)$.
Then, $Q_{12}(x,y)$ is given by~\cite{Raidal:2017mfl,Ballesteros:2018swv}
\begin{align}
	Q_{12}(x,y)
	&=
	Q_1(x)Q_2(y|x)
	=
	\bar n^2 (1+\xi(x)) (1+\xi(y))
	\exp\left[
	    -\Gamma(y)
	\right]
	\theta(y-x).
\end{align}
Using this formula, we can calculate the probability for PBHs to form with the initial separations $(x,y)$.

Once the initial configuration of three PBHs is given, we can check whether PBHs (I) and (II) form the binary.
We impose two conditions for the configuration of PBHs to result in the binary merger: PBH (I) and (II) should be bounded gravitationally overcoming the Hubble expansion; and PBH (III) should not fall into the binary.
The binary formation occurs if $t_{\rm ff}(r,z) <H(z)^{-1} $, where $t_{\rm ff}(r,z)$ is the free-fall time of the binary with the total mass $2M_{\rm PBH}$ with the initial separation $r$ at redshift $z$:
\begin{align}
    t_{\rm ff}(r,z)\equiv (2M_{\rm PBH} G)^{-1/2} \left( \frac{r}{1+z} \right)^{3/2}.
\end{align}
Since the Hubble time changes as  $H^{-1}\propto (1+z)^{-3/2}$ during the matter dominated era and $H^{-1}\propto (1+z)^{-2}$ during the radiation dominated era,
the ratio of $t_{\rm ff}(r,z)$ to $H(z)^{-1} $ decreases during the radiation dominated era and is constant during the matter dominated era.
Thus, the binary formation can occur only before the matter-radiation equality, and no more binaries are formed in the matter dominated era.
During the radiation dominated era, we approximate the energy density as $\rho(z)\simeq \rho_{c,0}\Omega_m (1+z)^4/(1+z_{\rm eq})$.
We define the ``decoupling time'' as $t_{\rm ff}(r,z_{\rm dec}) = H(z_{\rm dec})^{-1}$, and then $z_{\rm dec}(r)$ is given by
\begin{align}
    \frac{1+z_{\rm dec}(r)}{1+z_{\rm eq}}
    =
    \left(
    \frac{r_{\rm max}}{r}
    \right)^3
    \quad,\quad
     r_{\rm max}
    \equiv
    \left(
    \frac{3}{4\pi}
    \frac{2M_{\rm PBH}}{ \rho_{c,0}\Omega_m}
    \right)^{1/3}
    =
    7.1\times 10^{-4} \, {\rm Mpc}
    \left(
    \frac{M_{\rm PBH}}{30 M_\odot}
    \frac{0.3}{ \Omega_m}
    \frac{0.9\times 10^{-29}\,{\rm g~cm}^{-3}}{ \rho_{c,0}}
    \right)^{1/3}
    .
\end{align}
We require that the decoupling of PBHs (I) and (II) occurs during the radiation dominated era and that the decoupling of PBH (III) does not occur, which lead to the condition of the initial separation for the binary merger:
\begin{align}
    x <  r_{\rm max}  < y,
\end{align}
where we neglect an $\mathcal O(1)$ factor for PBH (III) due to a difference mass.

Once the binary is formed, it shrinks radiating gravitational waves and finally merges.
The coalescence time is determined by the initial radius and eccentricity of the binary.
The physical separation of PBHs (I) and (II) at $z_{\rm dec}(x)$ is given by
\begin{align}
    a(x)=\frac{x}{1+z_{\rm dec}}
    =
    \left(  \frac{x}{r_{\rm max}}\right)^3
    \frac{x}{1+z_{\rm eq}}.
\end{align}
The eccentricity of the binary $e$ is determined by the angular momentum induced by PBH (III).
When all the PBHs have the same mass, the eccentricity is roughly given by
\begin{align}
    e= \sqrt{1-
    \left( \frac{x}{y}\right)^6
    }.
\end{align}
Then, the coalescence time is given by
\begin{align}
    T(x,y) = R_m^{-3} a^4 (1-e^2)^{7/2}
    =\frac{x^{37}}{R_m^{3}r_{\rm max}^{12} y^{21}(1+z_{\rm eq})^4}
    \quad,\quad
    R_m=
    \left(\frac{3}{170}\right)^{-1/3}GM_{\rm PBH}
    =
    5.5\times 10^{-21} \, {\rm Mpc}
    ~\frac{M_{\rm PBH}}{30M_\odot}
    .
\end{align}
We can convert the probability distribution of the initial configuration $(x,y)$ into that of the coalescence time $t$ as
\begin{align}
    P_{\rm merge}(t)
    &=
    \int_0^{r_{\rm max}} 4\pi x^2\mathrm{d} x
    \int_{r_{\rm max}}^\infty 4\pi y^2\mathrm{d} y
    ~
    Q_{12}(x,y) ~
    \delta(t- T(x,y)) \nonumber
 \\    &=
    (4\pi)^2
    \int_0^{r_{\rm max}} x^2\mathrm{d} x
    ~
    y^2
    Q_{12}(x,y)
    \left(
    \frac{\mathrm{d} T(x,y)}{\mathrm{d} y}
    \right)^{-1}
    ~
    \theta(T(x,r_{\rm max})-t )
    \bigg |_{y=y_t(x)},
    \label{eq_Pmerge}
    \\
    y_t(x)
    &\equiv
    \left(
    \frac{x^{37}}{t R_m^{3}r_{\rm max}^{12}(1+z_{\rm eq})^4}
    \right)^{1/21},
\end{align}
where $y_t(x)$ represents the distance of PBH (III) to make the binary merge at $t = T(x,y_t(x))$.
The step function $\theta$ describes the condition $y>r_{\rm max}$, that is, $t=T(x,y)<T(x,r_{\rm max})$.
When PBH (III) is far away from the binary ($y\to \infty$), it exerts small torque on the binary ($e\to 1$), and the coalescence time becomes short ($T(x,y)\to 0$).
The step function, $\theta(T(x,r_{\rm max})-t )$, requires that the initial separation of the binary is large enough for the binary to survive until $t$, which determines the lower limit of integration as
\begin{align}
    x_t =
    \left(
    t R_m^{3}
    (1+z_{\rm eq})^4 r_{\rm max}^{33}
    \right)^{1/37}.
\end{align}

\begin{figure}[t]
	\centering
	\includegraphics[width=0.49\textwidth ]{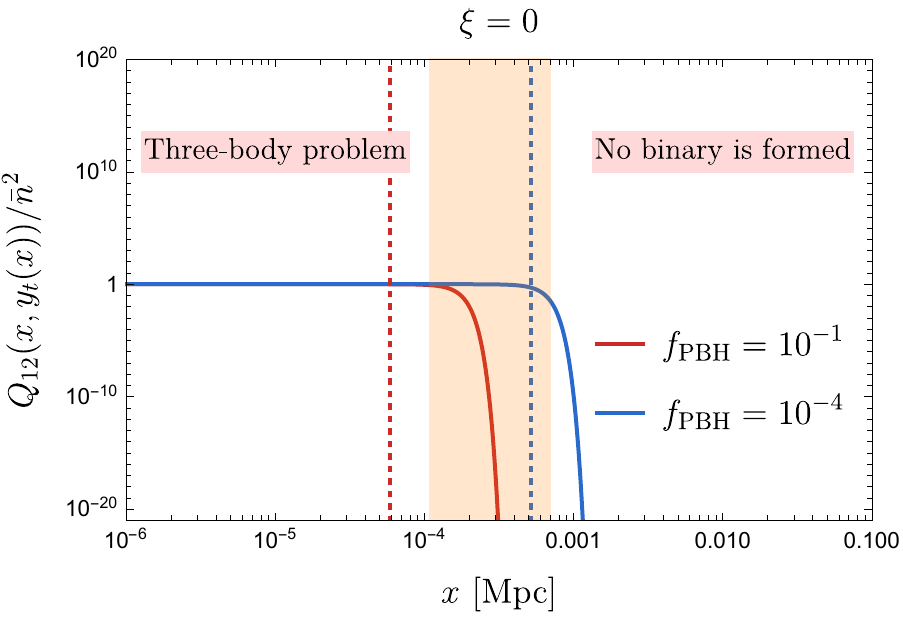}
	\includegraphics[width=0.49\textwidth ]{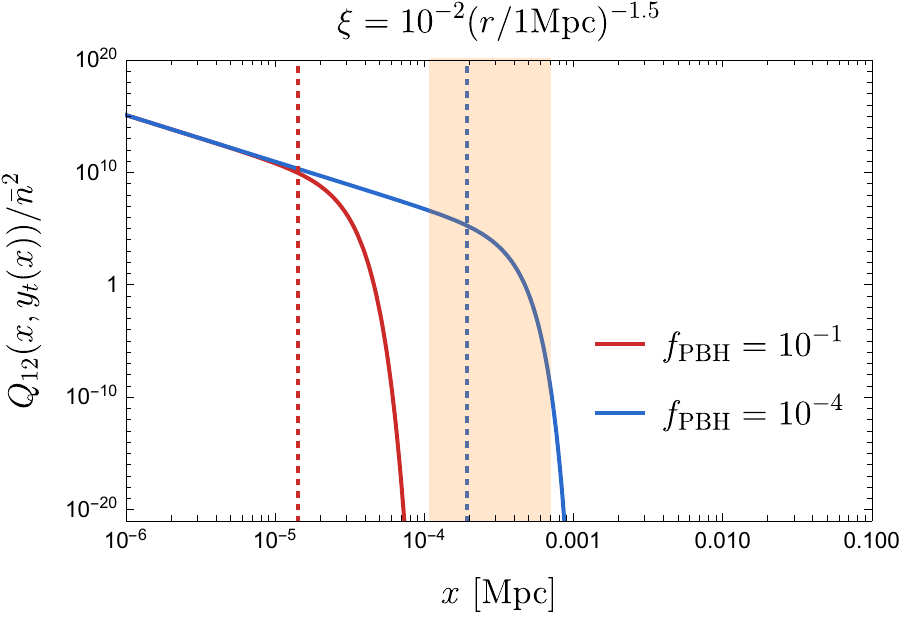}
	\caption{
	A part of integrand in the merger rate [Eq.\eqref{eq_Pmerge}] without (left) and with (right) clustering.
	The horizontal axes represent the separation of PBHs (I) and (II), and the position of PBH (III) is fixed so that the binary merges at $t = T(x,y_t(x))$.
	The color of lines represents the abundance of PBHs, $f_{\rm PBH} = 10^{-1}$ (red lines) and $10^{-4}$ (blue lines), respectively.
	The dotted lines represent $x$ satisfying $r_{\rm mean} = y_t(x)$, where the distance to the PBH (III) is equal to the mean separation of PBHs.
	The orange region is integration range, $x\in [x_t,r_{\rm max}]$.
	}
	\label{fig_integrand_Pmerge}
\end{figure}

In summary, the integration range of Eq.~\eqref{eq_Pmerge} is $x\in [x_t,r_{\rm max}]$.
The integrand of Eq.~\eqref{eq_Pmerge} is proportional to $x^{51/7}Q_{12}(x,y_t(x))$, and the integrated value depends on the functional form of  $Q_{12}(x,y_t(x))$.
We show $Q_{12}(x,y_t(x))$ in Fig.~\ref{fig_integrand_Pmerge}.
The orange region is the integration range, and PBH (III) is too close to avoid the three-body problem for $x<x_t$, and  PBHs (I) and (II) are too far to form the binary for $r_{\rm max}<x$.
The solid lines represent the normalized $Q_{12}(x,y_t(x))$ for PBH abundance $f_{\rm PBH} = 10^{-1}$ (red line) and $10^{-4}$ (blue line).
Without clustering (left panel), $Q_{12}(x,y_t(x))$ becomes $\bar n^2$ for small $x$.
For large $x$, the location of PBH (III), $y_t(x)$, should be far from the binary to merge at $t = T(x,y_t(x))$.
When the distance to PBH (III) becomes larger than the mean separation of PBHs, $y_t(x)> r_{\rm mean}$,
$\Gamma(y_t)$ suppresses $Q_{12}(x,y_t(x))$.
The dotted lines represent $x$ satisfying $r_{\rm mean} = y_t(x)$, and the suppression factor $\exp[-\Gamma(y_t)]$ becomes significant on the right side of the dotted lines.
With clustering (right panel), the suppression is stronger than the case without clustering, and the integrand is highly suppressed over the whole range of integration for $f_{\rm PBH} =10^{-1}$.
Thus, for a large PBH abundance, most of the binaries lead to the three-body problem with clustering.

\begin{figure}[th]
	\centering
	\includegraphics[width=0.8\textwidth ]{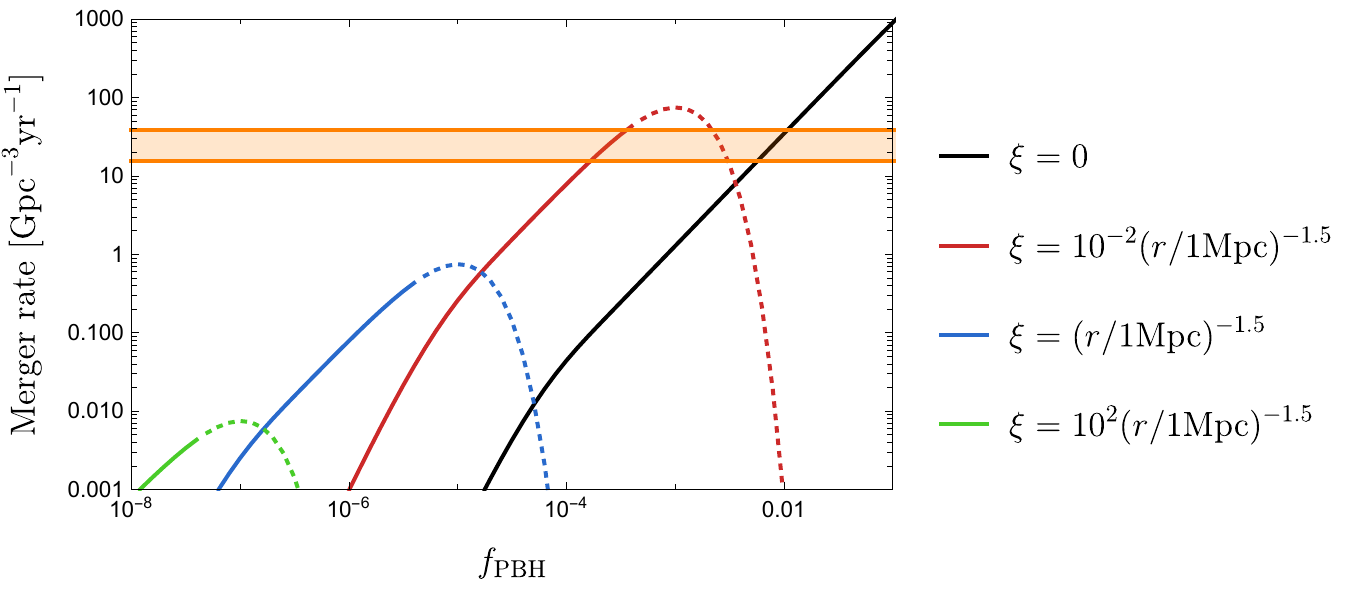}
	\caption{
	The merger rate distribution of PBHs with the mass 30~$M_\odot$ and the abundance $f_{\rm PBH}$.
	The orange line is the observed merger rate of LIGO-VIRGO experiment~\cite{Abbott:2020gyp}, $\mathcal R= 23.9^{+14.3}_{-8.6} ~{\rm Gpc}^{-3}{\rm yr}^{-1}$.
	The different lines represent the significance of the clustering with $\xi= \xi_* (r/1\mathrm{Mpc})^{-2}$.
	The dotted lines represent the condition $r_{\rm mean} < y_t(x_t) $,
	where PBHs are produced too close to avoid three-body problem.
	}
	\label{fig_example_Rmerge}
\end{figure}

The merger rate
at the current time $t_0 = 13.7$~Gyr is given by
\begin{align}
    \mathcal R_{\rm merger}
    =\bar n P_{\rm merge}(t_0)
    =
     8.2
     ~
     {\rm Gpc}^{-3}{\rm yr}^{-1}
    ~
    \frac{f_{\rm PBH}}{10^{-3}}
    \frac{t_0P_{\rm merge}(t_0)}{10^{-4}}
    \frac{30 M_\odot}{M_{\rm PBH}}
    \frac{ \Omega_{\rm DM}}{0.25}
    \frac{ \rho_{c,0}}{0.9\times 10^{-29}\,{\rm g~cm}^{-3}},
\end{align}
where $t_0 P_{\rm merge}(t_0)$ depends on the mass, abundance, and clustering of PBHs.
We numerically calculate the merger rate with some choices of clustering, which is shown in Fig.~\ref{fig_example_Rmerge}.
Although the LIGO-Virgo collaboration observes the binary mergers up to $z\lesssim 1$ \cite{Abbott:2020niy}, we simply compare the current merger rate  with the observed value (orange region), $\mathcal R= 23.9^{+14.3}_{-8.6} ~{\rm Gpc}^{-3}{\rm yr}^{-1}$~\cite{Abbott:2020gyp}.
Without clustering (black line), the merger rate increases as the PBH abundance $f_{\rm PBH}$ increases.
On the other hand, with clustering, the merger rate is amplified for small $f_{\rm PBH}$, while it is highly suppressed for large $f_{\rm PBH}$.
Since we calculate the merger rate conservatively by removing the binary merger with three-body dynamics, the merger rate is highly suppressed for large $f_{\rm PBH}$.
As shown in Fig.~\ref{fig_integrand_Pmerge}, the suppression occurs when the integrand is suppressed over the whole integration range, which is given by the condition $r_{\rm mean} < y_t(x_t) $ and shown in Fig.~\ref{fig_example_Rmerge} as the dotted lines.

The clustered PBH distribution drastically changes the merger rate density.
When the PBH abundance is small, the clustering increases the merger rate due to the increased binary formation rate.
On the other hand, when the PBH abundance is large, the PBHs tend to be too close to avoid the three-body problem, and it leads to the strong suppression of the merger rate in our formulation.

In our PBH formation model, the typical clustering effect [Eq.~\eqref{eq_xi_summary}] is so large that most binaries result in the three-body problem.
Thus, to precisely estimate the merger rate in our model, it is necessary to include the mergers in three-body systems of PBHs~\cite{Vaskonen:2019jpv}.
It is very difficult to analytically solve the three-body system, and numerical calculation is required, which is out of the scope of this paper.

In the end, we comment on the assumption on the monochromatic mass spectrum in the above analysis.
When the mass spectrum is broad, the merger rate increases by a factor of $\mathcal O(1)$ as shown in Fig.~3 of Ref.~\cite{Raidal:2018bbj}.
The authors fix the total abundance of PBHs and assume the log-normal distribution, which leads to a larger number of small PBHs.
Such small PBHs contribute to the merger rate, while the three-body problem could be more serious in this case.
We expect that the clustering is more important than the extended mass spectrum for the merger rate.

\section{Discussion and Summary}
\label{sec: discussion}

The spatial distribution of PBHs is an important property since it can drastically change the phenomena related to PBHs.
Although models of PBH formation usually assume the random distribution of PBHs, some models predict the clustered PBH distribution, for example, the models with the non-Gaussian perturbations and the models using the false vacuum to produce the bubble-like objects.

In this paper, we have investigated the clustering of PBHs in the PBH formation model using the Affleck-Dine baryogenesis.
We have studied the stochastic dynamics of the scalar field during inflation and  derived the PBH formation rate.
We also estimated the two-point correlation function of PBHs, which characterizes the clustering of PBHs.
It was found that the formed PBHs show a strong clustering,
which is consistent with the result derived in a different method~\cite{Shinohara:2021psq}.
Finally, we derive the reduced PBH correlation function given by Eq.~\eqref{eq_xi_summary}.

Using the reduced PBH correlation function, we have investigated the effect of the clustering on two phenomena related to PBHs; the isocurvature fluctuations and the merger rate distribution.
First, the clustering sources the isocurvature perturbations of PBHs in addition to the Poisson fluctuations.
We have estimated the power spectrum of density fluctuation of PBHs and have put the upper bound on the PBH abundance and the significance of clustering by using the current isocurvature constraints from the Planck satellite as shown in Fig.~\ref{fig_const_xif2}.

Second, the clustering of PBHs can drastically change the binary formation rate of PBHs, and the resultant merger rate is also modified.
We have found that the merger rate increases with the clustering for a small PBH abundance due to the enhanced binary formation rate, while it decreases for a large PBH abundance since the three-body problem occurs more frequently for the clustered PBHs.
As a result, it was found that it is difficult for our model to explain the LIGO-Virgo event rate of binary mergers when we conservatively neglect the binary merger in three-body systems.
While the treatment of the three-body problem is difficult, Ref.~\cite{Vaskonen:2019jpv} suggests that the perturbed binaries by the three-body problem still contribute to the merger rate.
The authors estimate $\mathcal O(10^2)$ suppression of the merger rate due to the three-body problem (See Fig.~2 of Ref.~\cite{Vaskonen:2019jpv} ).
Thus, it might be still possible that the strongly clustered PBHs explain the observed merger rate when we carefully estimate the merger of perturbed binaries.
It is left for future work to see whether our model can explain the merger rate observed by LIGO-VIRGO collaboration by correctly including PBH mergers in three-body systems.

\section*{Acknowledgments}

We would like to thank Tomoya Kinugawa for fruitful discussions and productive comments.
This work is supported by the Grant-in-Aid for Scientific Research Fund of the JSPS  20H05851(M.\,K.), 21K03567(M.\,K.),
20J20248 (K.\,M.) and
19J21974 (H.\,N.).
M.\,K. and K.\,M. are supported by World Premier International Research Center Initiative (WPI Initiative), MEXT, Japan (M.\,K. and K.\,M.).
K.\,M. is supported by the Program of Excellence in Photon Science.
H.\,N. is supported by Advanced Leading Graduate Course for Photon Science.

\appendix
\section{Probability distribution in the massive case}
\label{App: Prob distro in massive}
In this appendix, we consider the time evolution of the probability distribution of the radial component of a complex scalar field under the mass potential and the quantum fluctuation.
In this situation, the probability distribution of the complex scalar field follows the Fokker-Planck equation
\begin{equation}
  \frac{\partial P(N,\phi)}{\partial N}
  =
  \sum_{i = 1,2} \frac{\partial}{\partial \phi_i}
  \left[
    \frac{\partial V(\phi)}{\partial \phi_i}\frac{P(N,\phi)}{3H_I^2} +
    \frac{H_I^2}{8\pi^2} \frac{\partial P(N,\phi)}{\partial \phi_i}
  \right],
\end{equation}
where we consider the Hubble mass term as the mass potential
\begin{equation}
    V(\phi) = c_I H^2 |\phi|^2,
\end{equation}
with a dimensionless positive parameter $c_I$.
This equation can be rewritten with the polar coordinates $(\varphi, \theta) \equiv (|\phi|, \mathrm{arg}\phi)$ as
\begin{equation}
    \frac{\partial P(N,\phi)}{\partial N}
    =
    c_I'
    \left(
        P(N,\phi)
        + \frac{\varphi}{2} \frac{\partial P(N,\phi)}{\partial \varphi}
    \right)
    +
    \frac{H_I^2}{8\pi^2}
    \left(
       \frac{\partial^2 P(N,\phi)}{\partial \varphi^2}
        + \frac{1}{\varphi} \frac{\partial P(N,\phi)}{\partial \varphi}
        + \frac{1}{\varphi} \frac{\partial^2 P(N,\phi)}{\partial \theta^2}
    \right),
\end{equation}
where $c'_I \equiv 4c_I/3$.
Since we are interested in the probability distribution of the radial component of $\phi$, we integrate $P$ over $\theta$ and obtain
\begin{equation}
    \bar{P}(N,\varphi) = \varphi \int_0^{2\pi} \mathrm{d}\theta \, P(N,\phi).
\end{equation}
Note that $\bar{P}(N,\varphi)$ satisfies
\begin{equation}
    \int_0^{\infty} \mathrm{d}\varphi \, \bar{P}(N,\varphi) = 1.
\end{equation}
Then the differential equation that $\bar{P}(N,\varphi)$ follows is
\begin{equation}
    \frac{\partial \bar{P}(N,\varphi)}{\partial N}
    =
    \frac{c_I'}{2}
    \left(
        \bar{P}(N,\varphi)
        + \varphi \frac{\partial \bar{P}(N,\varphi)}{\partial \varphi}
    \right)
    +
    \frac{H_I^2}{8\pi^2}
    \left(
       \frac{\partial^2 \bar{P}(N,\varphi)}{\partial \varphi^2}
        - \frac{1}{\varphi} \frac{\partial \bar{P}(N,\varphi)}{\partial \varphi}
        + \frac{\bar{P}(N,\varphi)}{\varphi^2}
    \right).
\end{equation}
By using a dimensionless variable, $\tilde{\varphi} = 2\pi\varphi/H_I$, this equation becomes
\begin{equation}
    \frac{\partial \tilde{P}(N,\tilde{\varphi})}{\partial N}
    =
    \frac{c_I'}{2}
    \left(
        \tilde{P}(N,\tilde{\varphi})
        + \tilde{\varphi} \frac{\partial \tilde{P}(N,\tilde{\varphi})}{\partial \tilde{\varphi}}
    \right)
    +
    \frac{1}{2}
    \left(
       \frac{\partial^2 \tilde{P}(N,\tilde{\varphi})}{\partial \tilde{\varphi}^2}
        - \frac{1}{\tilde{\varphi}} \frac{\partial \tilde{P}(N,\tilde{\varphi})}{\partial \tilde{\varphi}}
        + \frac{\tilde{P}(N,\tilde{\varphi})}{\tilde{\varphi}^2}
    \right),
\end{equation}
where $\tilde{P} \equiv H_I \bar{P}/(2\pi)$ is normalized so that
\begin{equation}
    \int_0^{\infty} \mathrm{d}\tilde{\varphi} \, \tilde{P}(N,\tilde{\varphi}) = 1.
\end{equation}
Here we define $Q(N,\Phi)$ satisfying
\begin{equation}
    \tilde{P}(N, \tilde{\varphi}) = e^{\frac{c_I'N}{2}} Q(N,e^{\frac{c_I'N}{2}}\tilde{\varphi}).
\end{equation}
Then $Q(N,\Phi)$ follows
\begin{equation}
     \frac{\partial Q(N,\Phi)}{\partial N}
     =
    \frac{e^{c_I'N}}{2}
    \left(
       \frac{\partial^2 Q(N,\Phi)}{\partial \Phi^2}
        - \frac{1}{\Phi} \frac{\partial Q(N,\Phi)}{\partial \Phi}
        + \frac{Q(N,\Phi)}{\Phi^2}
    \right).
\end{equation}
Next we use $M \equiv (e^{c_I' N}-1)/c_I'$ and rewrite the differential equation as
\begin{equation}
     \frac{\partial Q_M(M, \Phi)}{\partial M}
     =
    \frac{1}{2}
    \left(
       \frac{\partial^2 Q_M(M, \Phi)}{\partial \Phi^2}
        - \frac{1}{\Phi} \frac{\partial Q_M(M, \Phi)}{\partial \Phi}
        + \frac{Q_M(M, \Phi)}{\Phi^2}
    \right),
\end{equation}
where $Q_M(M, \Phi) \equiv Q(N(M), \Phi)$.
This is the same as the differential equation that $\tilde{P}(N, \tilde{\varphi})$ satisfies in the massless case, $c_I' = 0$.
Moreover $\tilde{P}$ and $Q_M$ are the same at $N=0$.
Therefore we can directly use the solution of $\tilde{P}$ in the massless case as the solution of $Q_M$ in the massive case.
In order to calculate $B_2$ in Eq.~\eqref{eq: B2}, we consider the initial condition of $P(N=0,\phi) = \delta^{(2)}(\phi-\phi_{\mathrm{init}})$.
For this initial condition, the solutions of $P(N, \phi)$ and $\tilde{P}(N, \tilde{\varphi})$ are
\begin{equation}
    P(N, \phi; \phi_{\mathrm{init}})|_{c_I' = 0}
    =
    \frac{1}{2\pi N} \left( \frac{2\pi}{H_I} \right)^2
    \exp \left[ -\left( \frac{2\pi}{H_I} \right)^2\frac{|\phi-\phi_{\mathrm{init}}|^2}{2N} \right],
\end{equation}
and
\begin{equation}
    \tilde{P}(N, \tilde{\varphi}; \tilde{\varphi}_{\mathrm{init}})|_{c_I' = 0}
    =
    \frac{\tilde{\varphi}}{2\pi N}
    \int_0^{2\pi} \mathrm{d}\theta \,
    \exp \left[ -\frac{|\tilde{\phi}-\tilde{\phi}_{\mathrm{init}}|^2}{2N} \right],
\end{equation}
where $|\tilde{\phi}-\tilde{\phi}_{\mathrm{init}}|^2 = \tilde{\varphi}^2 + \tilde{\varphi}_{\mathrm{init}}^2 - 2\tilde{\varphi} \tilde{\varphi}_{\mathrm{init}} \cos{\theta}$.
Therefore, for the massive case, the solution of $Q(M, \Phi)$ is
\begin{equation}
    Q_M(M, \Phi; \Phi_{\mathrm{init}})
    =
    \frac{\Phi}{2\pi M}
    \int_0^{2\pi} \mathrm{d}\theta \,
    \exp \left[ -\frac{\Phi^2+\Phi_{\mathrm{init}}^2-2\Phi \Phi_{\mathrm{init}}\cos \theta}{2M} \right],
\end{equation}
and the solution of $\tilde{P}(N, \tilde{\varphi})$ is
\begin{align}
    \tilde{P}(N, \tilde{\varphi}; \tilde{\varphi}_{\mathrm{init}})
    &=
    \frac{c_I' \tilde{\varphi}}{2\pi (e^{c_I'} - 1)}
    e^{c_I' N}
    \int_0^{2\pi} \mathrm{d}\theta \,
    \exp \left[
        - c_I'
        \frac
        {e^{c_I' N} \tilde{\varphi}^2
         + \tilde{\varphi}_{\mathrm{init}}^2
         - 2e^{c_I' N/2}\tilde{\varphi} \tilde{\varphi}_{\mathrm{init}}\cos \theta}
        {2(e^{c_I'N} - 1)}
    \right]
    \nonumber\\
    &=
    \frac{\tilde{\varphi}}{2\pi \tilde{\sigma}^2(N)}
    \int_0^{2\pi} \mathrm{d}\theta \,
    \exp \left[
        -\frac
        {\tilde{\varphi}^2
         + e^{-c_I' N} \tilde{\varphi}_{\mathrm{init}}^2
         - 2e^{-c_I' N/2}\tilde{\varphi} \tilde{\varphi}_{\mathrm{init}}\cos \theta}
        {2 \tilde{\sigma}^2(N)}
    \right]
    \nonumber\\
    &=
    \frac{\tilde{\varphi}}{\tilde{\sigma}^2(N)}
    I_0 \left(
        e^{\frac{-c_I' N}{2}}\frac{\tilde{\varphi} \tilde{\varphi}_{\mathrm{init}}}{\tilde{\sigma}^2(N)}
    \right)
    \exp \left[
        -\frac{\tilde{\varphi}^2 + e^{-c_I' N}\tilde{\varphi}_{\mathrm{init}}^2}
        {2 \tilde{\sigma}^2(N)}
    \right],
\end{align}
where $I_0(z)$ is the modified Bessel function of the first kind of order $0$, and $\tilde{\sigma}^2(N) \equiv (1 - e^{-c_I' N})/c_I'$.
As a special case,
\begin{equation}
    \tilde{P}(N, \tilde{\varphi}; \tilde{\varphi}_{\mathrm{init}} = 0)
    =
    \frac{\tilde{\varphi}}{\tilde{\sigma}^2(N)}
    \exp \left[
        -\frac{\tilde{\varphi}^2 }{2 \tilde{\sigma}^2(N)}
    \right].
\end{equation}
It is easy to see that these solutions go to those in the massless case in the limit of $c_I' \to 0$.

\section{Reduced PBH correlation function}
\label{app: reduced PBH correlation function}

In this appendix, we derive the relation between $\xi(x)$ in Eq.~\eqref{eq_def_xi_finite} and $\tilde{\xi}(x)$ in Eq.~\eqref{eq_def_tildexi}.
Our formulation of the probability distribution of the IR mode of the AD field can use only $\beta_2$ as the quantity that represents the two-point correlation of PBHs.
Therefore, we assume that the realized number density of PBHs at a distance $x$ from the observer is given by $(1 + \tilde{\xi}(x))\bar{n}$ in our Universe.
This assumption is equivalent to assume that a PBH is located at the position of the observer.
Since the mergers of BHs have been observed on a much smaller scale compared with the current horizon scale, this assumption can be justified as an approximation.
Under this assumption, the number of PBHs in the observable Universe is given by
\begin{equation}
    N_\mathrm{obs}
    \equiv
    1 + \bar{n} \int \mathrm{d} V \, \left( 1+\tilde{\xi}(x) \right),
\end{equation}
where $V$ is the volume of the observable Universe and we included the PBH at the observer, which is not counted by the integration over $V$.
Note that $ N_\mathrm{obs}/V$ is larger than $\bar n$ due to the assumption above.
Then we should normalize $\tilde{\xi}(x)$ to correctly represent the reduced PBH correlation function and define $\xi(x)$ as
\begin{align}
    &
    \left( 1+\tilde{\xi}(x) \right) \bar{n}
    =
    \left( 1+\xi(x) \right) \frac{N_\mathrm{obs}}{V}
    \nonumber\\
    \Leftrightarrow \, &
    1 + \xi(x)
    =
    \frac{1 + \tilde{\xi}(x)}{F}
    \left( 1 - \frac{1}{N_\mathrm{obs}} \right),
\end{align}
where $F \equiv \int \mathrm{d}V \, (1 + \tilde{\xi}(x))/V$.
Note that $\xi(x)$ defined here satisfies $\int \mathrm{d} V \, \xi(x) = -V/N_\mathrm{obs}$, which is also derived from the standard definition of $\xi(x)$ in Eq.~\eqref{eq_delta_twopoint} in the monochromatic mass case.
Since we are interested in $N_\mathrm{obs} \gg 1$, we use the following quantities as those to be compared with the observed quantities:
\begin{align}
    \bar{n}_\mathrm{obs}
    & \equiv
    \frac{N_\mathrm{obs}}{V}
    \simeq
    F \bar{n},
    \\
    1 + \xi(x)
    & \simeq
    \frac{1 + \tilde{\xi}(x)}{F}.
\end{align}
The parameters that we used in Figs.~\ref{fig_abundance} and \ref{fig-xi} lead to $F \lesssim 2$ and then we approximate $F \sim 1$ in the order estimation of the PBH abundance and correlation function.

\small
\bibliographystyle{apsrev4-1}
\bibliography{Ref}

\end{document}